\documentclass[hidelinks]{article}
\usepackage{arxiv}
\AtBeginDocument{%
  \providecommand\BibTeX{{%
    \normalfont B\kern-0.5em{\scshape i\kern-0.25em b}\kern-0.8em\TeX}}}
\usepackage{algorithm}
\usepackage{algpseudocode}
\usepackage{algorithmicx}
\usepackage{enumerate}
\usepackage{pgfplots}
\usepackage{subcaption}
\usepackage{soul}
\usepgfplotslibrary{groupplots}
\usepackage{amsmath,amssymb,amsfonts}
\usepackage{booktabs}

\usepackage{wrapfig}
\usepackage{lscape}
\usepackage{rotating}
\usepackage{graphicx}
\usepackage{caption}
\usepackage{underscore}
\usepackage{multirow}
\usepackage[T1]{fontenc}
\usepackage{makecell}
\usepackage{changepage}
\usepackage[inkscapelatex=false]{svg}

\usepackage{array}
\usepackage{float}
\usepackage{placeins}
\usepackage{stackengine}
\usepackage{url}
\usepackage{numprint}
\usepackage{siunitx}

\begin{document}
\title{Towards Harnessing the Power of LLMs for ABAC Policy Mining
}

\author{%
  More Aayush Babasaheb\\
  Indian Institute of Technology Kharagpur, India\\
  \texttt{aayush.more@kgpian.iitkgp.ac.in} 
   \And
  Shamik~Sural \\
  Indian Institute of Technology Kharagpur, India\\
  \texttt{shamik@cse.iitkgp.ac.in} 
}


\date{}

\renewcommand{\headeright}{}
\renewcommand{\undertitle}{}

\maketitle
\begin{abstract}
This paper presents an empirical investigation into the capabilities of Large Language Models (LLMs) to perform automated Attribute-based Access Control (ABAC) policy mining. While ABAC provides fine-grained, context-aware access management, the increasing number and complexity of access policies can make their formulation and evaluation rather challenging. To address the task of synthesizing concise yet accurate policies, we evaluate the performance of some of the state-of-the-art LLMs, specifically Google Gemini (Flash and Pro) and OpenAI ChatGPT, as potential policy mining engines. An experimental framework was developed in Python to generate randomized access data parameterized by varying numbers of subjects, objects, and initial policy sets. The baseline policy sets, which govern permission decisions between subjects and objects, serve as the ground truth for comparison. Each LLM-generated policy was evaluated against the baseline policy using standard performance metrics. The results indicate that LLMs can effectively infer compact and valid ABAC policies for small-scale scenarios. However, as the system size increases, characterized by higher numbers of subjects and objects, LLM outputs exhibit declining accuracy and precision, coupled with significant increase in the size of policy generated, which is beyond the optimal size. These findings highlight both the promise and limitations of current LLM architectures for scalable policy mining in access control domains. Future work will explore hybrid approaches that combine prompt optimization with classical rule mining algorithms to improve scalability and interpretability in complex ABAC environments.
\end{abstract}



\section{Introduction}
\label{sec:intro}

Access control mechanisms have seen significant advancements over the last few decades, driven by the growing demand for flexibility, scalability, and security in managing permissions in complex information systems. Traditional models, such as Discretionary Access Control (DAC) and Role-based Access Control (RBAC), have played a fundamental role in defining access policies. RBAC, in particular, became a foundational standard by simplifying administration through the association of permissions with roles rather than individual users  \cite{ausanka2001methods}\cite{sandhu1996role}. However, as systems become more dynamic, distributed, and data-centric, the static nature of RBAC encounters several challenges. In large enterprises, managing a vast and growing number of roles can lead to the problem of role explosion, undermining the administrative benefits of the model \cite{das2018policy}. This necessity has led to the emergence of Attribute-based Access Control (ABAC), a paradigm that determines access rights based on the evaluation of attributes \cite{hu2014guide}. In an ABAC system, access decisions are influenced by a combination of subject attributes (e.g., role, department, security clearance), object attributes (e.g., resource type, security level, data sensitivity), and environmental conditions (e.g., time of access, location, threat level).

The attributes of entities in ABAC are evaluated against a set of rules (collectively called an ABAC policy) to determine access permissions, allowing for a more nuanced, fine-grained, and context-aware control as compared to the traditional static models. With ABAC, organizations can define detailed policies that dynamically adapt to the needs of modern information systems \cite{hu2014guide}. However, the power of ABAC introduces a significant challenge - that of policy engineering. The process of defining, managing, analyzing, and optimizing the potentially massive and intricate set of rules is complex, costly, and error-prone \cite{das2018policy}. This policy engineering bottleneck has become one of the primary obstacles to the widespread adoption of ABAC. Research towards overcoming this roadblock has largely bifurcated into two strategies, namely, top-down policy generation and bottom-up policy mining \cite{das2018policy}. The top-down approach attempts to translate high-level business requirements or natural language policies into machine-enforceable rules \cite{masoud3}\cite{narouei2017topdown}\cite{ramkrishnannlacp}\cite{indrakshisacmat23}\cite{stolleraccesslogmining}\cite{das2018using}. Policy mining, on the other hand, seeks to automatically infer or discover policies bottom-up by analyzing existing authorization data, such as access logs or legacy configurations (e.g., Access Control Lists - ACLs). Our work focuses on the latter, a problem that has been extensively studied using algorithmic approaches \cite{xu2015mining, talukdar2017efficient}.

With recent advancements in artificial intelligence, Large Language Models (LLMs) such as Google Gemini and OpenAI ChatGPT, etc., have demonstrated impressive capabilities in tasks involving pattern discovery and logical reasoning. This follows a trend in applying machine learning to access control, which previously utilized techniques like deep learning for processing log data \cite{mocanu2015towards, aln-nss-survey2022, DBLP:conf/codaspy/NobiKHSS22}. These developments motivate an investigation into whether general purpose LLMs can assist in automating the policy mining task, specifically in deriving concise policy sets that accurately capture access requirements from authorization data. The ability of such models to generalize from examples and infer logical structures suggests their potential applicability in deriving valid and optimized ABAC policies. In our work, a controlled experimental setup was developed using data generated through Python scripts. The use of generated data is a necessary methodological step towards creating a benchmark, as the field of access control automation suffers from lack of public real-world datasets \cite{aln-nss-survey2022}. Our generated data is comprised of randomly defined subjects and objects, each with associated attributes, along with a rule set governing permissions, collectively representing an access control environment. These predefined rule sets serve as the reference standard for evaluating the performance of different LLM configurations.

The LLM models Gemini Flash, Gemini Pro, and ChatGPT were tested using multiple prompting techniques and data input strategies to assess their ability to accurately learn and reproduce the underlying access control logic. Performance of each model was evaluated based on five key metrics: \textit{accuracy} - representing the percentage of correctly classified access decisions, \textit{precision} - indicating the proportion of correctly granted access permissions, \textit{recall} - reflecting the proportion of actual permitted accesses correctly identified, \textit{F1-score} - providing the harmonic mean of precision and recall, and \textit{rule minimality} - measuring the size of the generated policy set relative to the baseline. Experimental results demonstrate that LLMs are capable of generating coherent and consistent ABAC policies for relatively smaller system configurations. However, as the number of subjects and objects increases, the models exhibit a gradual decline in accuracy, precision, and recall, along with a growth in the policy size. These findings provide valuable insights into the current capabilities and limitations of LLMs in access control policy generation, and highlight the critical, unaddressed challenges of achieving scalability in such systems.

The rest of the paper is organized as follows. In the next section, we introduce some of the preliminary concepts on ABAC and LLMs. The proposed methodology is described in detail in Section \ref{sec:methodology}. Section \ref{sec:experimentalsetup} elaborates on the experimental setup while the results of our experiments are included in Section \ref{sec:Results}. Related work is reviewed in Section \ref{sec:related}. Finally, Section \ref{sec:concl} concludes the paper and provides directions for future research.

\section{Preliminaries}
\label{sec:prelims}

This section introduces the fundamental concepts underlying the application of large language models in the context of attribute-based access control policy engineering. It begins with outlining the principles of the ABAC model, followed by
a discussion on the policy mining problem. Finally, we provide an overview of how LLMs, through their interpretation and generation capabilities, can be prompted to support or automate different aspects of access control.

\subsection{Attribute-based Access Control}
\label{subsec:prelims-abac}

ABAC is an abstract model in which authorization specification and enforcement is based on the characteristics of the entities involved in an access request, rather than solely on the identity of the requester \cite{singh2019managing}. 

\subsubsection{ABAC Components}
\label{subsec:prelims-abac_components}
ABAC, as specified in NIST SP 800-162 \cite{hu2014guide}, primarily involves attributes and policies, which are evaluated by functional components to render and enforce access decisions as described below.
\begin{itemize}
    \item \textbf{Attributes:} These are characteristics of various entities involved in an access attempt. They are typically represented as name-value pairs. The core types are:
        \begin{itemize}
            \item \textbf{Subject Attributes (SA):} Characteristics describing the user or entity requesting access (e.g., role = `Doctor', department = `Cardiology', clearance = `TopSecret').
            \item \textbf{Object Attributes (OA):} Characteristics describing the resource or data being accessed (e.g., type = `MedicalRecord', sensitivity = `High', owner = `HospitalA').
            \item \textbf{Environmental Attributes (EA):} Contextual information relevant to the access request (e.g., timeOfDay = `09:00', accessLocation = `InternalNetwork', deviceType = `Trusted').
        \end{itemize}
    \item \textbf{Policy ($\Pi$):} A set of rules or relationships that specify the conditions under which access should be granted or denied, based on the evaluation of subject, object, and environmental attributes against the requested operation. Further details are given in the next sub-section.
    \item \textbf{Functional Components:} While the internal mechanisms are abstracted out in the model, the key ABAC functions involve:
        \begin{itemize}
            \item \textbf{Policy Decision Point (PDP):} Evaluates the applicable policies against the attributes provided in an access request context to compute an authorization decision (Permit/Deny) \cite{hu2014guide}.
            \item \textbf{Policy Enforcement Point (PEP):} Enforces the decision rendered by the PDP, allowing or blocking the requested operation \cite{hu2014guide}.
        \end{itemize}
\end{itemize}
An authorization in ABAC, such as a \textit{Doctor} from the \textit{Cardiology} department being permitted to \textit{read} a \textit{MedicalRecord} with \textit{High} sensitivity during \textit{InternalNetwork} access, is determined by evaluating the subject’s attributes (e.g., \textit{role = Doctor}, \textit{department = Cardiology}), the object’s attributes (e.g., \textit{type = MedicalRecord}, \textit{sensitivity = High}), and the environmental attributes (e.g., \textit{accessLocation = InternalNetwork}) against a relevant rule in the policy $\Pi$.

\subsubsection{Policy and Rule Structure}
A policy $\Pi$ in ABAC is comprised of a collection of access rules. Each rule $r \in \Pi$ typically represents a conjunction of attribute conditions that must be met for the rule to apply. Formally, a rule $r$ can often be characterized as a 4-tuple: $\langle \textit{SC}, \textit{OC}, \textit{EC}, \textit{OP} \rangle$.

\begin{itemize}
    \item $SC$: A set of required subject attribute conditions (e.g., role = `Faculty' AND department = `Science').
    \item $OC$: A set of required object attribute conditions (e.g., type = `Journal').
    \item $EC$: A set of required environmental attribute conditions (e.g., time = `Open hours').
    \item $OP$: The specific operation(s) (e.g., `read', `write', `execute') governed by the rule. This can be considered a set of permissible operations.
\end{itemize}
An access request is typically represented as a triplet $\langle \textit{u}, \textit{o}, \textit{op} \rangle$, where the user $u$ seeks to perform operation $op$ on object $o$. The PDP resolves this request by evaluating the attributes of $u$, $o$, and the current environment against the rules in $\Pi$. By default (in a permit-based policy), access is granted if the request satisfies the attribute conditions of at least one rule in $\Pi$ that permits the operation $op$. Otherwise, access is denied. Note that, in ABAC, the terms \textit{subject} and \textit{user} are used interchangeably. Likewise, \textit{object} and \textit{resource} are used interchangeably. 

\subsubsection{Illustrative Example}
Consider a simplistic university library system where the following ABAC components govern access to resources.
\begin{itemize}
    \item SA: \{role, department\}
    \item OA: \{type\}
    \item EA: \{time\}
    \item OP: \{View, Borrow, Return\}
\end{itemize}
Each attribute can take discrete categorical values from the following domains:
\begin{itemize}
    \item $V^{S}_{role}$ = \{Student, Faculty, Staff\}
    \item $V^{S}_{department}$ = \{Science, Arts, Engineering\}
    \item $V^{O}_{type}$ = \{Book, Journal, Thesis\}
    \item $V^{E}_{time}$ = \{Open hours, Closed hours\}
\end{itemize}
The policy $\Pi$ for this system consists of the rules shown in Table \ref{tab:tab1}. For simplicity, rules use the `$=$' operator. A `*' denotes a wildcard matching any value.

\begin{table}[h!]
    \centering
    \caption{Sample ABAC Policy for University Library System}
    \label{tab:tab1}
    \setlength{\tabcolsep}{6pt}
    \renewcommand{\arraystretch}{1}
    \begin{tabular}{|c|c c c c|c|}
        \hline
        \textbf{Rule} & \textbf{Role} & \textbf{Department} & \textbf{Type} & \textbf{Time} & \textbf{Operation}\\
        \hline
        $r_1$ & Student & Science & Book & Open hours & View \\
        $r_2$ & Faculty & Science & Journal & Open hours & Borrow \\
        $r_3$ & Faculty & * & Thesis & * & View \\
        $r_4$ & Staff & Arts & Book & Open hours & Return \\
        \hline
    \end{tabular}
\end{table}

Based on these rules, if a user $u$ with attributes $\{$role=Faculty, department=Science$\}$ requests to perform the operation `Borrow' on resource $o$ with attribute $\{$type=Journal$\}$ during an environment where $\{$time=Open hours$\}$, the request $\langle u, o, \text{Borrow} \rangle$ satisfies the conditions of rule $r_2$. Access would therefore be granted by the PDP. This example illustrates how rules combine attribute conditions to define permissions and how wildcards provide generalization. Discovering such generalized rules from raw data is the essence of the policy mining task.

\subsection{Policy Engineering in ABAC}
\label{subsec:prelim-policy_engineering}

As introduced earlier, the flexibility of ABAC leads to a significant challenge in policy engineering: the design, creation, management, and optimization of ABAC policies \cite{das2018policy}. This process is critical but often complex and error prone when performed manually \cite{das2018policy}. Formally, the problem of policy mining can be defined as follows \cite{talukdar2017efficient, xu2015mining}:
\textit{Given a set of users $U$, a set of objects $O$, sets of user attributes $SA$ and object attributes $OA$, sets $S_v$ and $O_v$ containing all user/object attribute-value assignments, and a set of known authorizations $A$ (representing permitted $\langle u, o, op \rangle$ tuples, potentially derived from logs or an existing access control matrix), derive an ABAC policy $\Pi$ such that:}

\begin{enumerate}
    \item \textit{$\Pi$ is consistent with $A$: The set of accesses permitted by $\Pi$ is exactly the set $A$.}
    \item \textit{The number of rules in $\Pi$ is minimized (or optimized according to some complexity metric).}
\end{enumerate}
Environmental attributes ($E_a$, $E_v$) can also be included in the input if available from logs or context.
As outlined by Das et al. \cite{das2018policy}, policy engineering encompasses several strategies:
\begin{itemize}
    \item \textbf{Top-Down Approaches:} Policies are derived from high-level organizational requirements, business process descriptions, or Natural Language Policy (NLP) documents \cite{das2018policy}. While potentially ensuring alignment with business intent, these methods are often manual, resource-intensive, and may struggle with ambiguity \cite{das2018policy}. Efforts exist to automate parts of this process using NLP techniques \cite{narouei2017topdown}.
    \item \textbf{Bottom-Up Approaches (Policy Mining):} Policies are inferred algorithmically from existing authorization data, as defined above \cite{das2018policy}. This is the focus of the our work. Key research includes developing efficient algorithms \cite{talukdar2017efficient, das2018using}, mining under practical constraints \cite{gautam_et_al}, and utilizing machine learning techniques like deep learning from logs \cite{mocanu2015towards}. The foundational algorithm by Xu and Stoller \cite{xu2015mining} demonstrated mining from ACLs and attribute data.
    \item \textbf{Hybrid Approaches:} These aim to combine the strengths of both top-down and bottom-up methods, leveraging data while incorporating semantic or business process knowledge to produce more meaningful and accurate policies \cite{das2018hype}.
\end{itemize}

A major sub-problem within policy engineering as noted above, and the focus of this work, is \textit{policy mining}. Policy mining aims to automate the creation of ABAC policies by analyzing existing authorization evidence, such as system configurations, access logs, or permissions derived from legacy models like DAC and RBAC \cite{xu2015mining, talukdar2017efficient, das2018policy}. This bottom-up approach is particularly useful for migrating to ABAC or for discovering the implicit rules governing current access patterns \cite{das2018policy}. It can help produce optimized policies by identifying redundancies or overly permissive rules \cite{talukdar2017efficient}. It has previously been shown that Policy Mining is an NP-Complete problem. Hence, the available solutions attempt to develop heuristic or randomized algorithms that generate close to optimal solutions in polynomial time. In this paper, we investigate the potential of LLMs to perform the bottom-up policy mining task, specifically focusing on their ability to inductively reason from authorization data for producing minimal yet correct rule sets.

\subsection{Large Language Models and Prompt Engineering}
\label{subsubsec:prelims-llms}

Large language models represent a class of machine learning models characterized by their massive scale (billions of parameters) and training on vast amounts of text and code data. This extensive pre-training endows them with powerful capabilities for natural language understanding, generation, pattern discovery, and logical reasoning. Recent models have shown surprising aptitude for tasks requiring complex, step-by-step reasoning, even in domains beyond natural language. These developments motivate the investigation undertaken in this paper - exploring whether LLMs can effectively assist in automating complex, logic-based tasks such as ABAC policy mining. Their inherent ability to generalize from examples and infer logical structures from data suggests a potential applicability for generating valid, concise, and optimized ABAC policies directly from authorization sets \cite{sivasothy2024large}. 


It is widely known that the performance of LLMs is highly sensitive to how a desired task is presented to them. The input text provided to the model, known as the prompt, serves as the instruction and context. Prompt Engineering is the process of designing, refining, and structuring these input prompts to effectively guide the LLM towards generating the desired output accurately and in the correct format. The experimental methodology employed in this paper utilizes several established prompting strategies to assess their impact on the policy mining task:
\begin{itemize}
    \item \textbf{Zero-Shot Prompting:} The model receives a detailed description of the task, input format, and any constraints, but is given \textit{no} illustrative examples. It must perform the task based solely on its pre-trained understanding and the provided instructions. 
    \item \textbf{Few-Shot Prompting:} The prompt includes not only the task description but also a small number of examples demonstrating the expected input-output behavior. These examples help the model understand the desired logic, format, and level of generalization. 
    \item \textbf{Chain-of-Thought (CoT) Prompting:} This is a more elaborate technique where the prompt explicitly instructs the model to articulate its reasoning process step-by-step before arriving at the final answer. Often, the prompt includes an example demonstrating this intermediate reasoning (e.g., parsing data, identifying patterns, generalizing rules, checking against negative examples, minimizing the set). CoT aims to elicit more robust and reliable reasoning from the LLM, particularly for complex tasks.
\end{itemize}
Understanding these techniques is crucial for interpreting the experimental results presented later in this paper, which systematically evaluate the impact of different prompting strategies on the ability of LLMs to perform ABAC policy mining.

\section{Proposed Methodology}
\label{sec:methodology}
This section delineates our comprehensive methodology and a meticulously designed experimental framework employed to investigate the capabilities of LLMs in the domain of ABAC policy mining. As mentioned before, the primary objective of our work is to empirically evaluate the efficacy of prominent LLMs in synthesizing concise, accurate, and minimal access control policies from a given set of authorization decisions. The section begins by formalizing the representation of access control decisions, the different access data representation format possible, and what type of prompts, constraints as well as policy types that could be tested with LLMs.

\subsection{Formal Representation of Access Control Decisions}
\label{subsec:acm_theory}

In any given ABAC system, an access request is typically characterized by a triplet $(u, o, e)$, where a user or subject $u \in U$ attempts to perform an operation on an object $o \in O$ within a specific environment $e \in E$. The sets $U$, $O$, and $E$ represent the complete set of subjects, objects, and environmental conditions within the system, respectively. The ABAC policy decision point (PDP) evaluates a set of rules against the attributes of the subject, object, and environment to grant or deny this request. To systematically represent all possible authorizations in a finite system, we can conceptualize a three-dimensional Access Control Matrix (ACM). Each element in this matrix, indexed by a specific $(u, o, e)$ tuple, holds a binary value representing the outcome of an access request. A value of `1' signifies that access is granted (a permit decision), while a value of `0' indicates that access is denied (a deny decision). This ACM, therefore, serves as a complete and unambiguous representation of the protection state for every possible interaction in the system under consideration. It functions as the ground truth against which the policies generated by the LLMs are to be evaluated.

For the scope of this investigation, the framework was simplified by focusing exclusively on subject and object attributes, thereby omitting the environmental dimension. This simplification results in a two-dimensional ACM, where each \textit{cell} $(u, o)$ represents the access decision for a specific subject-object pair. Such a reduction was deemed appropriate as the fundamental challenge of policy mining, i.e., discovering logical relationships among attributes to explain access decisions, remains intact. The principles and methodologies developed for a two-dimensional system can be readily generalized to incorporate environmental attributes, which can be treated as additional attribute-value pairs associated with each access request. This focused approach allows for a more controlled and interpretable analysis of LLM performance on the core task of rule inferencing.

\subsection{Data Input Formats}
\label{subsec:input_format}

The structure and organization of input data plays a critical role in determining how effectively an LLM can interpret access control relationships. In the context of policy mining, the same underlying access control information can be represented in multiple ways—each emphasizing different aspects of the subject, object, and decision relationships. To illustrate these formats, we use a \( 2 \times 2 \) ACM, where two subjects (\( S_1, S_2 \)) and two objects (\( O_1, O_2 \)) are defined.

\subsubsection*{Sample Data}

\begin{table}[H]
\centering
\caption{Sample Access Control Matrix (ACM)}
\begin{tabular}{c|cc}
\toprule
\textbf{Subjects / Objects} & \textbf{O\textsubscript{1}} & \textbf{O\textsubscript{2}} \\
\midrule
\textbf{S\textsubscript{1}} & 1 & 0 \\
\textbf{S\textsubscript{2}} & 0 & 1 \\
\bottomrule
\end{tabular}
\end{table}

Each subject and object is associated with two uniformly formatted attribute-value pairs, as shown below:

\begin{table}[H]
\centering
\caption{Subject and Object Attributes}
\begin{tabular}{lclcl}
\toprule
\textbf{Entity} & & \textbf{Attribute Values} \\
\midrule
S\textsubscript{1} & : & SA\textsubscript{1}=Dept\_A,  SA\textsubscript{2}=Role\_X & \\
S\textsubscript{2} & : & SA\textsubscript{1}=Dept\_B,  SA\textsubscript{2}=Role\_Y & \\
O\textsubscript{1} & : & OA\textsubscript{1}=Type\_A,  OA\textsubscript{2}=Level\_1 & \\
O\textsubscript{2} & : & OA\textsubscript{1}=Type\_B,  OA\textsubscript{2}=Level\_2 & \\
\bottomrule
\end{tabular}
\end{table}

The following sub-sections describe how this same data is represented under the three evaluated input methods.

\subsubsection{ACM + Attributes (ACM Input Method)}
\label{subsubsec:ACM_method1}

In this format, the complete ACM is presented as a structured matrix, followed by separate lists of subject and object attributes. This structure explicitly separates access decisions from entity characteristics, requiring the model to learn the mapping between them.

\begin{verbatim}
ACM:
        O1  O2
S1      1   0
S2      0   1

Subject Attributes:
S1: SA1=Dept_A, SA2=Role_X
S2: SA1=Dept_B, SA2=Role_Y

Object Attributes:
O1: OA1=Type_A, OA2=Level_1
O2: OA1=Type_B, OA2=Level_2
\end{verbatim}

\subsubsection{Access Data Logs (Access Data Input Method)}
\label{subsubsec:Access_method2}

This flattened representation merges all subject and object attributes for each ACM entry into a single line, along with the corresponding decision value. Each record effectively represents one cell of the ACM and provides the model with a direct mapping between attribute combinations and access decisions.

\begin{verbatim}
S1: SA1=Dept_A, SA2=Role_X | O1: OA1=Type_A, OA2=Level_1 | Decision=1
S1: SA1=Dept_A, SA2=Role_X | O2: OA1=Type_B, OA2=Level_2 | Decision=0
S2: SA1=Dept_B, SA2=Role_Y | O1: OA1=Type_A, OA2=Level_1 | Decision=0
S2: SA1=Dept_B, SA2=Role_Y | O2: OA1=Type_B, OA2=Level_2 | Decision=1
\end{verbatim}

\subsubsection{ACL + Attributes (ACL Input Method)}
\label{subsubsec:ACL_method}

In this format, the ACM is replaced by an Access Control List (ACL), which holds, for each object, the subjects permitted to access it. This method includes only positive (permit) authorizations, alongside the full attribute descriptions of all subjects and objects. It evaluates the model’s ability to infer rules from permission-focused data.

\begin{verbatim}
Access Control Lists:
O1: [S1]
O2: [S2]

Subject Attributes:
S1: SA1=Dept_A, SA2=Role_X
S2: SA1=Dept_B, SA2=Role_Y

Object Attributes:
O1: OA1=Type_A, OA2=Level_1
O2: OA1=Type_B, OA2=Level_2
\end{verbatim}

\noindent
Through these three representations, our study investigates how different data organization strategies affect the ability of the LLM to understand access control relationships and generalize policy rules.

\subsection{Prompts, Constraints and Policy Types used in the Methodology}
\label{subsec:prompt_method}

\begin{enumerate}[i]
    \item \textbf{Zero-Shot (No Examples):} This prompt provides a detailed description of the task, input format, and the critical safety constraints. However, it includes no illustrative examples, forcing the model to rely entirely on its understanding of the instructions. 
    
    \item \textbf{Example-Based:} This prompt provides the same task description and constraints as the zero-shot variant but includes several illustrative examples showing the input data and the corresponding correct minimal output. Two configurations are used: one \{\textit{without reasoning}\}, 
    where only the examples are shown, and another \{\textit{with reasoning}\} 
    , where each example also includes an explanation of the intermediate steps leading to the output.
    
    \item \textbf{Negative Constraint (No 0 to 1):} This prompt explicitly instructs the model not to generate any rule that would convert a `0' in the original ACM to a `1' in the reconstructed ACM. This \textit{no false positives} constraint tests the model’s ability to adhere to strict negative security requirements. 
    
    \item \textbf{Permit and Deny Policies (Deny Allowed):} This prompt allows the model to generate both \textit{permit} and \textit{deny} rules. In cases where a subject–object pair matches both a permit and a deny rule, a deny-overrides conflict resolution strategy is applied. This means that if any applicable rule denies access, the final decision is `deny', regardless of any matching permit rules. This setup mimics the behavior of real-world access control systems and tests the model’s ability to reason about conflicting policy statements. 
    
    \item \textbf{Chain of Thought (CoT):} This prompt explicitly instructs the model to follow a step-by-step reasoning process. It includes an example of the thought process, guiding the model to parse and filter, find patterns, validate generalizations against `deny` entries, and minimize the rules before producing the final output. 
\end{enumerate}

Details of the above prompts are given in the Appendix.

\section{Experimental Setup}
\label{sec:experimentalsetup}

This section details the complete configuration of the experimental setup used to assess the performance of LLMs in mining ABAC policies. It outlines the process of generating controlled datasets, the criteria used to evaluate the quality of mined policies, and the phased experimental design adopted to progressively analyze model behavior under varying configurations. Together, these components establish a systematic foundation for fair comparison and meaningful interpretation of results in the next section.

\subsection{Data Generation}
\label{subsec:synth_data}

To ensure a controlled and reproducible experimental environment, a data generation pipeline was developed using Python. This approach was essential because suitable real-world ABAC data is difficult to find. Even when such data is available, it presents several research challenges. First, they do not have sufficient variability across specific parameters (e.g., number of attributes, policy complexity) to test scalability, and it is not possible to know the definitive, optimal ground-truth policy that generated the access decisions. This lack of a known correct answer makes it difficult to reliably evaluate a mined solution.

Our data pipeline directly addresses these limitations by creating a diverse range of ABAC scenarios, including subjects, objects, their associated attributes, and, most importantly, the known ground-truth Access Control Matrix. The generation process is governed by several key parameters that allow for systematic variation in the complexity and characteristics of the policy mining problem. This provides a reliable foundation for evaluating mined policies against a known solution, ensuring a controlled and reproducible experimental environment.

\subsubsection{Generation of a Consistent Access Control Matrix}
\label{subsubsec:ACM-generation}

A critical requirement for a valid experimental setup is the generation of a consistent ACM. Consistency, in this context, implies adherence to the fundamental principle of ABAC, i.e., subjects (or objects) with identical attribute sets should exhibit identical access permissions. If two subjects, $u_1$ and $u_2$, possess the same attribute-value pairs, then for any given object $o$, the access decision for $(u_1, o)$ must be the same as that for $(u_2, o)$. A randomly populated ACM would not, in general, satisfy this consistency constraint.

To address this requirement, our data generator does not populate the ACM directly. Instead, it begins by creating an initial randomized set of ground-truth ABAC rules. These rules dictate the conditions under which access is granted. Each rule is a conjunction of attribute-value pair specifications for subjects and objects. For example, a rule might state: ``Permit access if a subject has attribute \texttt{SA\_1 = S\_1\_A} and the object has attribute \texttt{OA\_2 = O\_2\_B}.''
Once this initial rule set is established, the ACM is populated by systematically evaluating every subject-object pair $(u, o)$ against this set. If a pair's attributes satisfy at least one rule, the corresponding cell in the ACM is set to `1'; otherwise, it is set to `0'. This rule-based generation process guarantees that the resulting ACM is internally consistent and reflects an underlying logical policy.

\subsubsection{Experimental Parameters}
\label{subsubsec:exp_parameters}

The data generation process is controlled by the following parameters, which were systematically varied to assess LLM performance under different conditions.
\begin{enumerate}[i]
    \item \textbf{ACM Size:} The dimensions of the ACM were varied by changing the number of subjects ($|U|$) and the number of objects ($|O|$). This parameter directly controls the scale of the problem.
    \item \textbf{ACM Density:} Here, density refers to the proportion of `1's in the ACM. It was controlled by adjusting the number and specificity of the rules in the initial, ground-truth rule set. ACMs in real-world systems are often sparse (low density of `1's), making this an important parameter for simulating realistic scenarios.
    \item \textbf{Attribute Space:} The number of subject and object attributes, along with the cardinality of their respective value sets, were configurable parameters. This configurability enabled controlled variation in the complexity of the underlying access control logic and facilitated regulation of the ACM density.
\end{enumerate}

\subsubsection{Initial Rule Set as an Upper Bound on the Optimal Policy Size}

The set of rules used to generate the ACM serves a dual purpose. It not only ensures consistency as explained above, but also acts as an upper bound on the size of the optimal solution to the policy mining problem for that specific ACM. Since this rule set is known to perfectly describe all the authorizations in the ACM, its size (i.e., the number of rules) represents a target for rule minimality. ABAC policy mining, being an NP-Complete problem, there is no other way of efficiently determining such an upper bound on the optimal policy size. The performance of the LLMs can thus be evaluated not only on their ability to accurately reproduce the ACM but also on how closely the size of their generated policy set approaches (or if it is even less than) the size of this known, initial rule set.

\subsection{Evaluation Criteria}
\label{subsec:eval_crit}

To comprehensively evaluate the policies generated by LLMs, we used a set of standard performance metrics commonly applied in machine learning and information retrieval literature. These metrics were chosen to suit the access control domain, especially for addressing the data imbalance often seen in sparse access control matrices. In the discussions, a True Positive (respectively, True Negative) denotes that a Permit (respectively Deny) entry in the ground-truth ACM, is also obtained as a Permit (respectively, Deny) entry in the re-constructed ACM. False Negatives and False Positives are obtained from True Positives and True Negatives using their standard definitions.  

After an LLM generates a set of rules, this new rule set is used to reconstruct a new ACM. The reconstructed matrix is then compared element-by-element with the original, ground-truth ACM to calculate the following metrics.
\begin{itemize}
    \item \textbf{Policy Set Size:} The number of rules in the generated policy set. A primary goal of policy mining is to find a minimal set of rules. The mined policy is compared against the original, ground-truth rule set. This can be expressed as a ratio of the policy set size generated by LLM to the size of ground-truth rule set.

    \item \textbf{Accuracy:} This metric measures the overall correctness of the generated policy. It is calculated as the proportion of all access decisions (both permit and deny) that were correctly determined by the LLM generated rules.
    \begin{equation*}
        \text{Accuracy} = \frac{\text{True Positives} + \text{True Negatives}}{\text{Total Decisions}}
    \end{equation*}
    
    \item \textbf{Precision:} Precision measures the exactness of the permit decisions. It is the proportion of granted access requests in the reconstructed ACM that were also granted in the original ACM. High precision indicates that the generated policy does not grant unauthorized access.
    \begin{equation*}
        \text{Precision} = \frac{\text{True Positives}}{\text{True Positives} + \text{False Positives}}
    \end{equation*}

    \item \textbf{Recall:} Recall measures the completeness of the permit decisions. It is the proportion of all permit decisions from the original ACM that were successfully identified by the generated policy. High recall indicates that the policy does not improperly deny legitimate access requests.
    \begin{equation*}
        \text{Recall} = \frac{\text{True Positives}}{\text{True Positives} + \text{False Negatives}}
    \end{equation*}
    
    \item \textbf{F1-Score:} The F1-Score is the harmonic mean of Precision and Recall. It provides a single, balanced measure of a policy's performance, which is particularly useful when dealing with imbalanced datasets, such as sparse ACMs.
    \begin{equation*}
        \text{F1-Score} = 2 \times \frac{\text{Precision} \times \text{Recall}}{\text{Precision} + \text{Recall}}
    \end{equation*}
\end{itemize}

The use of all five metrics is crucial. A smaller policy set indicates a more concise and generalizable representation of the access control logic, reflecting the ability of the model to minimize redundancy while preserving correctness. For a sparse ACM, a model that generates a policy resulting in all `0's would achieve very high accuracy but with zero recall, rendering it useless. Conversely, a policy that permits all access requests would achieve perfect recall but suffer from extremely low precision and poor accuracy. The F1-Score provides a necessary trade-off, offering a more holistic assessment of the model's ability to correctly identify the permit decisions, which are often the primary focus of policy mining. Even small deviations in accuracy can have significant operational implications. For instance, a mere 1\% drop in accuracy on an ACM with 10,000 elements would require approximately 100 additional exception-handling rules to maintain the soundness and completeness of the system, imposing a considerable administrative burden. Thus, higher score on the latter four metrics may need to be given more importance than minimal size of the policy set.

\subsection{Phased Experiments}
\label{subsec:phases_exp}

The experimental investigation was structured into a series of phases, beginning with a broad evaluation of multiple models and progressively narrowing down to a more detailed analysis of the most promising candidates and techniques. For the first three phases, the dataset used consisted of a moderately sized ACM containing 15 subjects and 15 objects. This size was determined empirically to represent an effective middle ground: smaller ACMs (e.g., 5$\times$5) were overly simplistic, leading all models to perform near-perfectly, and thus obscuring performance differences. In contrast, substantially larger ACMs were computationally intensive, which could confound the comparative analysis. The 15$\times$15 configuration provides sufficient complexity to highlight model disparities while remaining computationally manageable.

Experiments were conducted by varying the density of `1's in the ACM to observe how performance changed as the number of positive examples increased. The same ACM input data was used across the first three phases to ensure consistency in evaluation. The analysis was conducted on a dataset comprised of five distinct test cases (TC1--TC5), each representing a different ACM density (i.e., percentage of `1' or permit entries). These test cases, described in Section~\ref{subsec:synth_data}, were generated for a 15$\times$15 ACM (225 total decisions) using a ground-truth policy set of ten rules, which served as the optimal benchmark for evaluating the \textit{Policy Set Size} metric. The specific densities for the five test cases were as follows: TC1 (29 `1's, 12.88\%), TC2 (69 `1's, 30.67\%), TC3 (91 `1's, 40.55\%), TC4 (104 `1's, 46.22\%), and TC5 (118 `1's, 52.44\%).

Unless otherwise specified, the input mode for the models was \textit{access data}. This configuration was chosen because combining attribute information with permission values enables the model to more effectively associate attribute combinations with their corresponding access outcomes, compared to presenting them separately in alternative input formats. To minimize the effect of random variation in model outputs and to prevent anomalous results, a regeneration protocol was applied. If an initial result from any model was deemed anomalous---defined as achieving an accuracy or precision below 0.9, or producing a non-minimal policy set where the number of rules equaled the number of `permit' entries (i.e., a trivial solution)---the prompt was regenerated. The better performing of the two results was retained for final analysis.


The initial phase aimed to identify the best-performing models from a selection of state-of-the-art LLMs: Google's Gemini 2.5 Flash and Gemini 2.5 Pro, and OpenAI's ChatGPT. Throughout this paper, these models will be referred to by their shorter names: Gemini Flash, Gemini Pro, and ChatGPT, respectively. In this phase, the models were tasked with generating only \textit{permit} rules that exclusively grant access. This simplified the task, allowing for a clear comparison of their core rule inference capabilities.


The second phase investigated the impact of different data input structures on model performance. The three formats evaluated—ACM + Attributes, Access Data Logs, and ACL + Attributes, were described in detail in Sub-section \ref{subsec:input_format}. Only the best method identified was retained for further analysis in the Scalability Analysis phase (Last phase).


The third phase analyzed the effect of various prompting strategies and policy constraints. The specific prompts—including Zero-Shot, Example-Based, Negative Constraint, Permit and Deny, and Chain of Thought—were fully detailed in Sub-section \ref{subsec:prompt_method}. Only the best two prompts were retained for further analysis in the next phase.

In the final phase, the insights gathered from the first three phases were made use of. The best-performing model, prompting strategies, and data input format were combined into an optimized configuration. This configuration was then systematically tested on ACMs of increasing size to evaluate the scalability of the LLM-based policy mining approach. Such an analysis is crucial for understanding the practical viability of using LLMs in larger, more realistic enterprise-level access control environments. The results and findings of these experiments are presented in the next section.

\section{Experimental Results}
\label{sec:Results}

This section provides a comprehensive analysis of the experimental results obtained from the setup described in Section \ref{sec:experimentalsetup}. It systematically presents and interprets the outcomes across multiple phases of experimentation, as outlined there. The analysis aims to provide a clear understanding of how different configurations and constraints influence the overall performance of policy mining. In all the graphs and tables, the policy set size is represented as a ratio of the policy size generated by the LLM to the size of the corresponding ground-truth policy for ease of understanding of the efficacy of using LLMs in policy mining. Unless otherwise mentioned, by `optimal benchmark', we mean the size of the initial policy that serves as an upper bound on the size of the optimal policy for the given problem instance.

\subsection{Phase 1: Initial Model Screening}
\label{subsec:phase1_screening}
We first study the capabilities of different LLMs for the policy mining task, before moving on to deeper investigations. 
\begin{figure}[t]
  \centering
  \includegraphics[width=\linewidth]{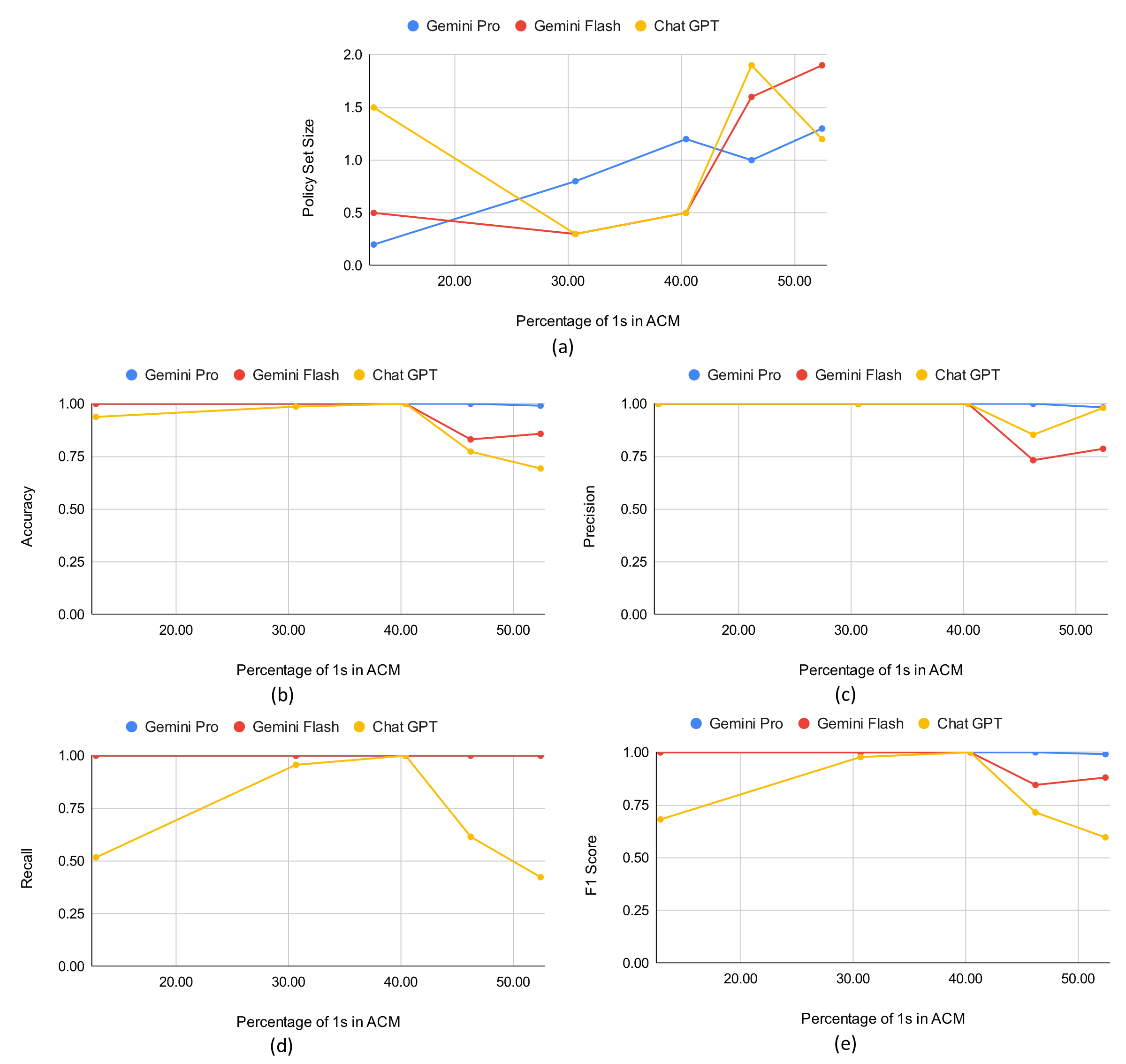}
  \caption{Comparison of LLM Models}
  \label{fig:model_com}
\end{figure}
The results of this initial screening as visualized in Figure \ref{fig:model_com}, reveal a consistent trend: the size of the generated policy set for all models generally increases with the density of the ACM (i.e., the percentage of `1's). While some models occasionally produce policy sets smaller than the 10-rule benchmark, the majority of the generated sets exceed this minimal size.
It is also observed that model performance varies significantly. Gemini 2.5 Pro demonstrates excellent capability, achieving near-perfect scores across all evaluation metrics, indicating strong potential for this task. Conversely, ChatGPT faces considerable difficulty, yielding poor scores on TC1, TC4, and TC5. Gemini 2.5 Flash delivers a moderate and stable performance, maintaining perfect recall across all test cases and achieving accuracy and F1-scores consistently above 0.8. Given these findings, Gemini 2.5 Pro and Gemini 2.5 Flash are selected for the subsequent phases of our investigation. ChatGPT is excluded from further study.

\subsection{Phase 2: Investigation on Data Input Formats}
\label{subsec:phase2_input_formats}

This phase evaluates the impact of different data input formats, described in Sub-section \ref{subsec:input_format}, on individual model performance.
\begin{figure}[t]
  \centering
  \includegraphics[width=\linewidth]{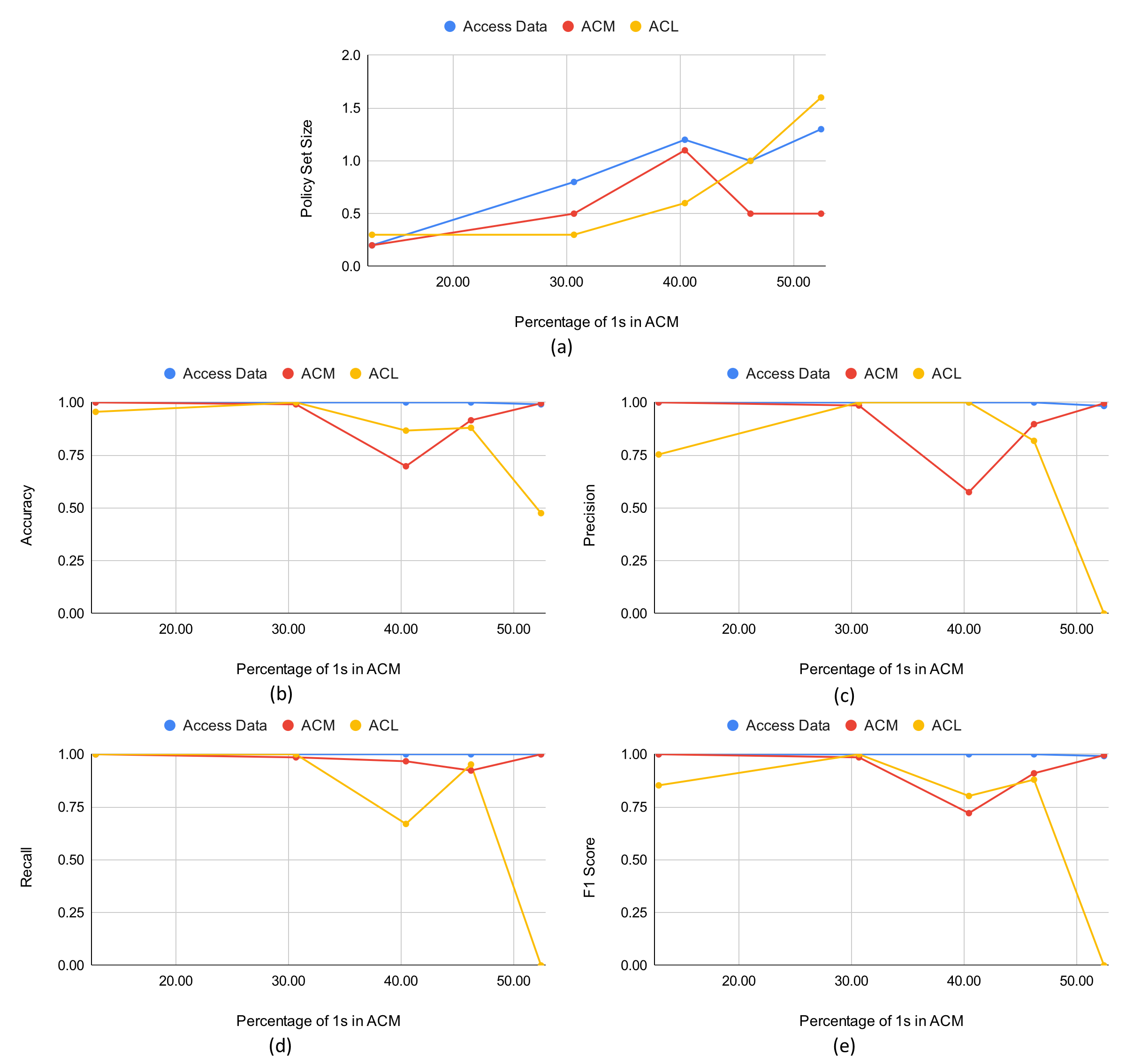}
  \caption{Comparison of Gemini Pro across Data Input Methods}
  \label{fig:input_comp_pro}
\end{figure}
As illustrated in Figure \ref{fig:input_comp_pro}, the performance of Gemini Pro varies with the input method. The \textit{Access Data} input method is superior, achieving near-perfect scores on all evaluation metrics across all test cases. The \textit{ACM} input method yields lower than expected results on TC3 and TC4 with respect to accuracy, precision and F1 Score, despite producing the most minimal policy sets for TC4 and TC5. The \textit{ACL} input method demonstrates inconsistent and erratic performance, with significantly poor evaluation scores on TC3 and TC5.
\begin{figure}[t]
  \centering
  \includegraphics[width=\linewidth]{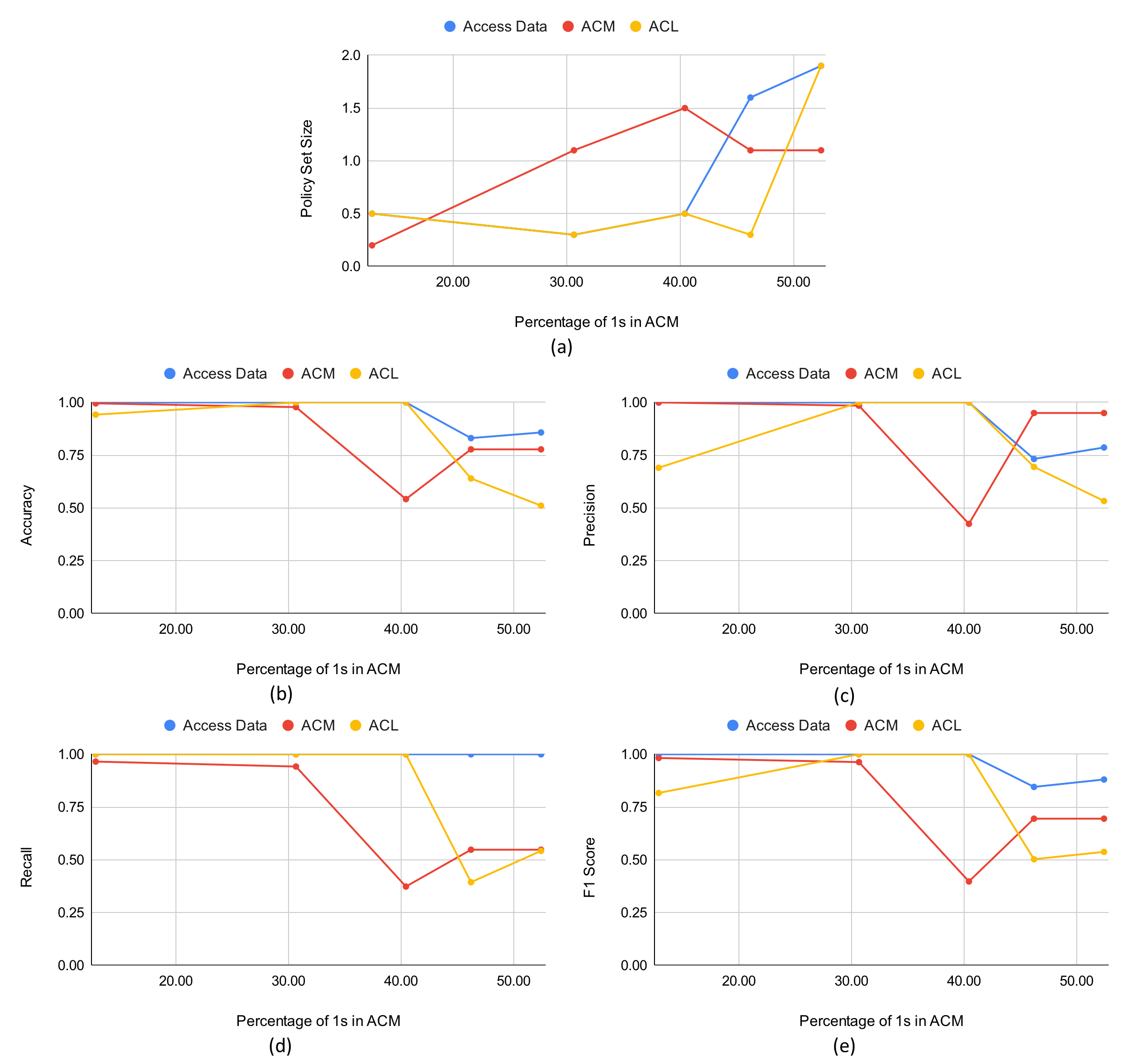}
  \caption{Comparison of Gemini Flash across Data Input Methods}
  \label{fig:input_comp_flash}
\end{figure}
The analysis for Gemini Flash, shown in Figure \ref{fig:input_comp_flash}, mirrors these findings. The \textit{Access Data} format provides the most stable performance, achieving perfect scores on the first three test cases and maintaining a perfect recall and an F1-Score above 0.8 for the remaining two. In contrast, the \textit{ACM} input method results in poor evaluation scores for TC3, TC4, and TC5. The \textit{ACL} input method performs well on the first three test cases but fails on TC4 and TC5.

Given its consistent and superior performance across both models, the \textit{Access Data} input format is identified as the most robust input strategy. Consequently, it is the only input method selected for the remaining phases.

\subsection{Phase 3: Analysis of Prompts, Constraints and Policy Types}
\label{subsec:phase3_prompt}

We next analyze the performance of Gemini Pro and Gemini Flash when subjected to the different prompt strategies outlined in Sub-section \ref{subsec:prompt_method}. It uses \textit{Access Data} input format, based on the findings of the previous section.
For Gemini Pro, the results are presented in Tables \ref{tab:prompt_rul_pro} through \ref{tab:prompt_f1_pro}. Graphical representations are omitted for this model, as its performance is exceptionally high and nearly identical across most configurations, with evaluation metrics consistently approaching 1.0. \textit{Prompt 2} and \textit{Prompt 3} are not tested on Gemini Pro, given the model’s strong ability to generalize from limited context. Consequently, its performance on these two prompts is expected to be comparable to that on \textit{Prompt 1}, owing to the similar nature of task descriptions across the prompts. For Gemini Pro, the \textit{No 0 to 1 Prompt} is quite effective at rule minimization, generating small policy sets (Table \ref{tab:prompt_rul_pro}), and at times achieving a size smaller than the optimal baseline, all while maintaining near-perfect evaluation metrics. \textit{Prompt 1} also performs strongly, with near-perfect metrics and policy set sizes comparable to the baseline. The \textit{Deny Allowed Prompt} is a notable exception, performing poorly on evaluation metrics for TC1 and TC5 and producing a significantly larger policy set for TC4. The \textit{COT Prompt} performs moderately well, with compact policy sets and high-metric scores, faltering only on TC1.

\begin{table}[htbp]
  \centering
  \caption{Comparison of Gemini Pro Policy Sets across Prompt Types}
  \label{tab:prompt_rul_pro}
  \begin{tabular}{lrrrr}
    \toprule
    Percentage of 1s & Prompt 1 & Deny Allowed & No 0 to 1 & COT \\
    \midrule
    12.88 & 0.2 & 1.1 & 0.2 & 0.2 \\
    30.67 & 0.8 & 0.5 & 0.7 & 0.8 \\
    40.44 & 1.2 & 0.7 & 1.2 & 0.7 \\
    46.22 & 1.0 & 2.2 & 0.6 & 1.0 \\
    52.44 & 1.3 & 1.0 & 0.6 & 0.4 \\
    \bottomrule
  \end{tabular}
\end{table}

\begin{table}[htbp]
  \centering
  \caption{Comparison of Gemini Pro Accuracy across Prompt Types}
  \label{tab:prompt_acc_pro}
  \begin{tabular}{lrrrr}
    \toprule
    Percentage of 1s & Prompt 1 & Deny Allowed & No 0 to 1 & COT \\
    \midrule
    12.88 & 1.00 & 0.87 & 1.00 & 0.85 \\
    30.67 & 1.00 & 1.00 & 1.00 & 1.00 \\
    40.44 & 1.00 & 1.00 & 1.00 & 0.99 \\
    46.22 & 1.00 & 1.00 & 1.00 & 0.98 \\
    52.44 & 0.99 & 0.45 & 1.00 & 1.00 \\
    \bottomrule
  \end{tabular}
\end{table}
  
\begin{table}[htbp]
  \centering
  \caption{Comparison of Gemini Pro Precision across Prompt Types}
  \label{tab:prompt_pre_pro}
  \begin{tabular}{lrrrr}
    \toprule
    Percentage of 1s & Prompt 1 & Deny Allowed & No 0 to 1 & COT \\
    \midrule
    12.88 & 1.00 & 0.00 & 1.00 & 0.47 \\
    30.67 & 1.00 & 1.00 & 1.00 & 1.00 \\
    40.44 & 1.00 & 1.00 & 0.99 & 0.97 \\
    46.22 & 1.00 & 1.00 & 1.00 & 0.96 \\
    52.44 & 0.98 & 0.42 & 1.00 & 1.00 \\
    \bottomrule
  \end{tabular}
\end{table}

\begin{table}[htbp]
  \centering
  \caption{Comparison of Gemini Pro Recall across Prompt Types}
  \label{tab:prompt_rec_pro}
  \begin{tabular}{lrrrr}
    \toprule
    Percentage of 1s & Prompt 1 & Deny Allowed & No 0 to 1 & COT \\
    \midrule
    12.88 & 1.00 & 0.00 & 1.00 & 1.00 \\
    30.67 & 1.00 & 1.00 & 1.00 & 1.00 \\
    40.44 & 1.00 & 1.00 & 1.00 & 1.00 \\
    46.22 & 1.00 & 1.00 & 1.00 & 1.00 \\
    52.44 & 1.00 & 0.12 & 1.00 & 1.00 \\
    \bottomrule
  \end{tabular}
\end{table}
  
\begin{table}[htbp]
  \centering
  \caption{Comparison of Gemini Pro F1 Score across Prompt Types}
  \label{tab:prompt_f1_pro}
  \begin{tabular}{lrrrr}
    \toprule
    Percentage of 1s & Prompt 1 & Deny Allowed & No 0 to 1 & COT \\
    \midrule
    12.88 & 1.00 & 0.00 & 1.00 & 0.64 \\
    30.67 & 1.00 & 1.00 & 1.00 & 1.00 \\
    40.44 & 1.00 & 1.00 & 0.99 & 0.98 \\
    46.22 & 1.00 & 1.00 & 1.00 & 0.98 \\
    52.44 & 0.99 & 0.19 & 1.00 & 1.00 \\
    \bottomrule
  \end{tabular}
\end{table}

\begin{figure}[t]
  \centering
  \includegraphics[width=\linewidth]{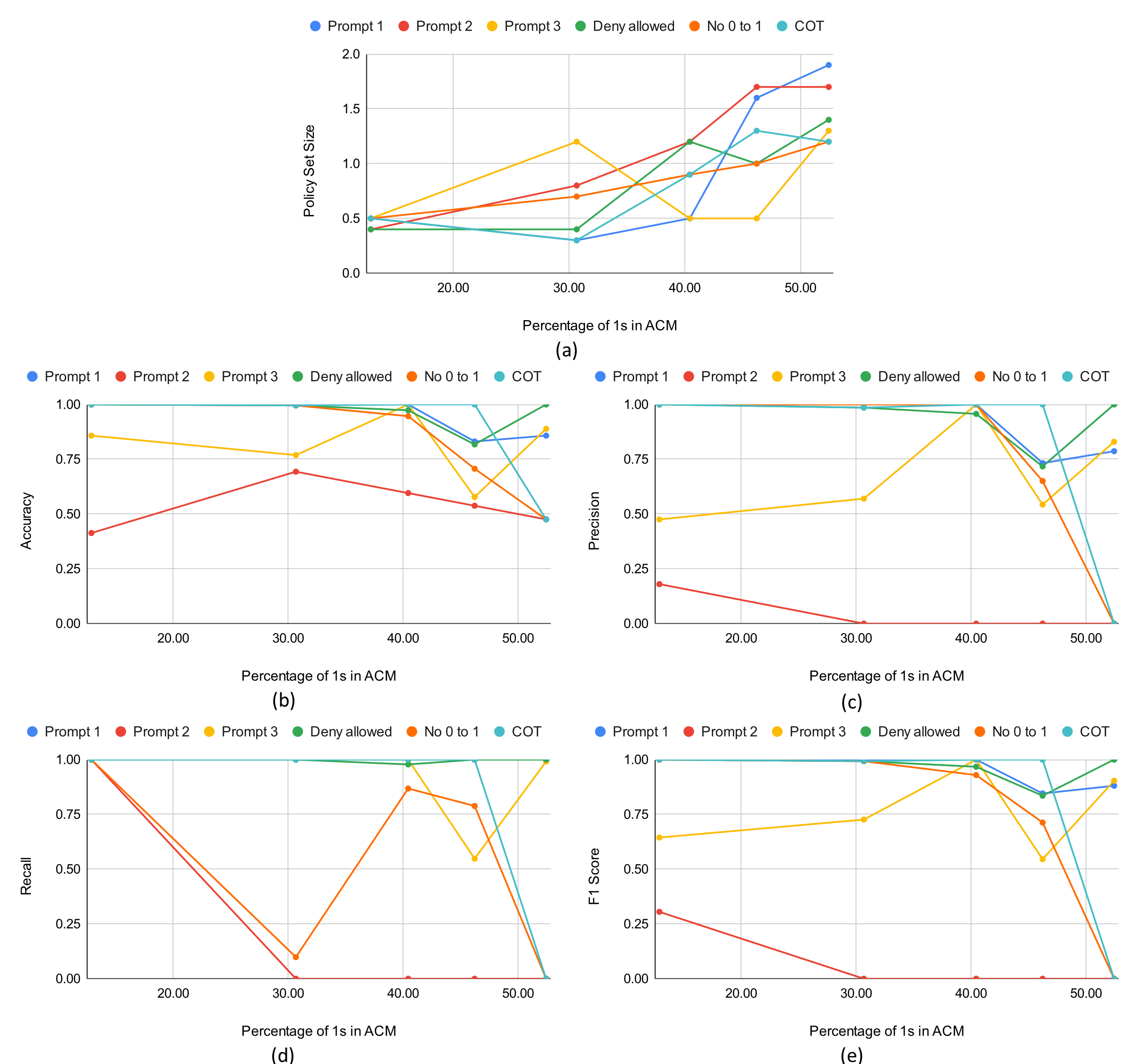}
  \caption{Comparison of Gemini Flash across Prompt Types } 
  \label{fig:prompt_comp_flash}
\end{figure}

The performance of Gemini Flash, as depicted in Figure \ref{fig:prompt_comp_flash}, shows greater variability across different prompt strategies. \textit{Prompt 1} and the \textit{Deny Allowed Prompt} both demonstrate relatively stable performance. Their accuracy and F1-scores remain above 0.8, with recall scores at or near 1.0 for all test cases. \textit{Prompt 2} (Zero-Shot) and \textit{Prompt 3} (Example-Based, No Reasoning) perform exceptionally poorly. The absence of illustrative examples or explicit reasoning guidance render them incapable of producing a useful policy set. The \textit{No 0 to 1 Prompt}, while producing relatively small policy sets, exhibit a clear declining trend in accuracy, precision, and F1-score, coupled with erratic recall performance. The \textit{Chain of Thoughts (COT) Prompt} achieves near-perfect scores across all metrics for TC1 through TC4, but experience a significant performance collapse on TC5.

For the scalability analysis in the next phase, two prompt strategies are selected. The primary criterion for selection is the correctness of the generated policy (i.e., the evaluation metrics) rather than the minimality (Size of Policy Set). This decision is based on the premise that a lapse in accuracy and similar metrics is a more severe failure in an access control system. 
Based on this criterion, \textit{Prompt 2} and \textit{Prompt 3} are immediately excluded due to their severe failure. The \textit{No 0 to 1 Prompt} is also excluded due to its clear declining performance trend on Gemini Flash. \textit{Prompt 1} is selected for its consistent and stable performance across both models. Finally, a choice is made between the \textit{Deny Allowed Prompt} and the \textit{Chain of Thoughts (COT) Prompt}. The \textit{Deny Allowed Prompt} exhibits poor performance on Gemini Pro (TC1, TC5), whereas the \textit{COT Prompt} performed poorly only on Gemini Flash (TC5). Given its superior performance on the more capable Gemini Pro model, the \textit{Chain of Thoughts (COT) Prompt} is selected as the second prompt for the final scalability analysis.

\subsection{Phase 4: Scalability Analysis}
\label{subsec:phase4_scalability}

Based on the outcomes of the preceding phases, the final phase of our investigation focuses on scalability analysis. The objective of this phase is to evaluate the performance of the selected models and prompt strategies when applied to ACMs of progressively increasing size. For this purpose, we choose the four most promising configurations identified in earlier experiments: Gemini Pro (Prompt 1), Gemini Flash (Prompt 1), Gemini Pro (Chain of Thought \& CoT), and Gemini Flash (Chain of Thought \& CoT). Since all of these configurations utilize Access Data as the input method, based on the results of Sub-section~\ref{subsec:phase2_input_formats}, this detail will henceforth be omitted for brevity and be assumed throughout this section.

The experimental setup is modified to generate ACMs with total elements ranging from 500 to 5000. Under the assumption that ACMs are typically sparse, the density of `1's is maintained at approximately 10\% for all test cases. Furthermore, the initial ground-truth policy set used to generate these ACMs is fixed at 20 rules. This figure serves as the new optimal benchmark for the `Policy Set Size' metric throughout the scalability analysis.

\subsubsection{Rule Minimality}
\label{subsubsec:ruleminimality}

The first metric evaluated is the size of the generated policy set, with results presented in Table~\ref{tab:scalability_rules}. The data reveals considerable instability and challenges in maintaining rule minimality as the system scales. The benchmark for an optimal policy set is 20 rules. No single configuration demonstrates consistently stable performance. Gemini Pro (COT) often produces the most compact rule sets (e.g., ratios of 0.2 at 1000 elements and 0.1 at 1500 elements), indicating fewer than the 20-rule benchmark. However, it expands substantially at 4000 elements (ratio 1.7). Gemini Flash (COT) is particularly volatile: it achieves a low ratio of 0.3 at 1000 elements but spikes dramatically to 5.6 at 4000 elements, indicating excessive policy generation. The Prompt 1 configurations are similarly erratic. Gemini Pro (Prompt 1) achieves excellent minimality at 1500 elements (0.1 ratio) but performs poorly at 700 and 2000 elements (1.1 ratio) and fails entirely at 5000 elements ($*$). This trend shows that while the models \textit{can} infer a minimal policy, they fail to do so reliably as problem size increases.

\begin{table}[htbp]
  \centering
  \caption{Scalability Analysis: Policy Set Size}
  \label{tab:scalability_rules}
  \begin{tabular}{@{}lcccc@{}}
    \toprule
    Total Elements in ACM & \begin{tabular}[c]{@{}c@{}}Gemini Flash\\ (COT)\end{tabular} & \begin{tabular}[c]{@{}c@{}}Gemini Pro\\ (COT)\end{tabular} & \begin{tabular}[c]{@{}c@{}}Gemini Flash\\ (Prompt 1)\end{tabular} & \begin{tabular}[c]{@{}c@{}}Gemini Pro\\ (Prompt 1)\end{tabular} \\
    \midrule
500 & 0.9 & 0.4 & 0.4 & 0.5 \\
700 & 1.3 & 0.5 & 0.6 & 1.1 \\
1000 & 0.3 & 0.2 & 0.8 & 0.3 \\
1500 & 0.8 & 0.1 & 0.7 & 0.1 \\
2000 & 0.6 & 0.5 & 0.2 & 1.1 \\
3000 & 1.3 & 0.5 & 1.6 & 0.4 \\
4000 & 5.6 & 1.7 & 0.7 & 1.7 \\
5000 & 0.5 & 0.3 & 0.3 & $*$ \\
\bottomrule
  \end{tabular}
  \vspace{1mm}
  
  \footnotesize{$*$ indicates that the model failed to produce a valid output.}
\end{table}

\subsubsection{Policy Correctness (Accuracy, Precision, Recall, F1 Score)}

The evaluation of policy correctness, detailed in Tables~\ref{tab:scalability_accuracy} through~\ref{tab:scalability_f1}, reveals increasing volatility and a marked decline in performance across all models as ACM size increases. Table~\ref{tab:scalability_accuracy} shows that Accuracy, while often appearing high (an expected artifact of sparse ACMs), fluctuates sharply. For example, Gemini Pro (Prompt 1) drops from 0.96 at 500 elements to 0.76 at 700 and further down to 0.64 at 4000 elements. Similarly, Gemini Flash (Prompt 1) maintains poor accuracy at small scales (0.76 at both 500 and 700 elements). Even the COT configurations show degradation, with Gemini Pro (COT) declining from near-perfect accuracy at 1000 elements to 0.67 at 4000 elements.

\begin{table}[htbp]
  \centering
  \caption{Scalability Analysis: Accuracy}
  \label{tab:scalability_accuracy}
  \begin{tabular}{@{}lcccc@{}}
    \toprule
    Total Elements in ACM & \begin{tabular}[c]{@{}c@{}}Gemini Flash\\ (COT)\end{tabular} & \begin{tabular}[c]{@{}c@{}}Gemini Pro\\ (COT)\end{tabular} & \begin{tabular}[c]{@{}c@{}}Gemini Flash\\ (Prompt 1)\end{tabular} & \begin{tabular}[c]{@{}c@{}}Gemini Pro\\ (Prompt 1)\end{tabular} \\
    \midrule
500 & 0.96 & 0.99 & 0.76 & 0.96 \\
700 & 1.00 & 0.99 & 0.76 & 0.76 \\
1000 & 1.00 & 1.00 & 1.00 & 1.00 \\
1500 & 0.77 & 0.99 & 0.89 & 0.67 \\
2000 & 1.00 & 0.92 & 0.96 & 0.92 \\
3000 & 0.96 & 0.81 & 0.90 & 0.85 \\
4000 & 0.98 & 0.67 & 0.91 & 0.64 \\
5000 & 0.89 & 0.94 & 0.80 & $*$ \\
\bottomrule
  \end{tabular}
  \vspace{1mm}
  
  \footnotesize{$*$ indicates that the model failed to produce a valid output.}
\end{table}

Precision and Recall (Tables~\ref{tab:scalability_precision} and~\ref{tab:scalability_recall}) provide a clearer picture of model behavior. Precision highlights each model’s ability to avoid false positives, which is critical for access control systems. Here, the Prompt 1 configurations consistently fail: Gemini Flash (Prompt 1) shows catastrophic precision scores of 0.26 at 700 elements and 0.20 at 5000 elements, while Gemini Pro (Prompt 1) falls to 0.17 at 4000 elements. Even the COT models struggle at larger scales, with Gemini Pro (COT) collapsing to 0.23 at 4000 elements. 

Recall, which measures the ability to capture valid “permit” cases, is equally unstable. Gemini Flash (COT) maintains near-perfect recall up to 2000 elements but drops drastically to 0.15 at 5000 elements. Gemini Flash (Prompt 1) also fails at 3000 elements (0.12 recall), missing most legitimate access permissions.

\begin{table}[htbp]
  \centering
  \caption{Scalability Analysis: Precision}
  \label{tab:scalability_precision}
  \begin{tabular}{@{}lcccc@{}}
    \toprule
    Total Elements in ACM & \begin{tabular}[c]{@{}c@{}}Gemini Flash\\ (COT)\end{tabular} & \begin{tabular}[c]{@{}c@{}}Gemini Pro\\ (COT)\end{tabular} & \begin{tabular}[c]{@{}c@{}}Gemini Flash\\ (Prompt 1)\end{tabular} & \begin{tabular}[c]{@{}c@{}}Gemini Pro\\ (Prompt 1)\end{tabular} \\
    \midrule
500 & 0.75 & 0.90 & 0.31 & 0.76 \\
700 & 1.00 & 0.91 & 0.26 & 0.33 \\
1000 & 1.00 & 1.00 & 1.00 & 1.00 \\
1500 & 0.29 & 1.00 & 0.49 & 0.24 \\
2000 & 1.00 & 0.54 & 1.00 & 0.55 \\
3000 & 0.94 & 0.36 & 0.76 & 0.33 \\
4000 & 0.99 & 0.23 & 0.54 & 0.17 \\
5000 & 0.61 & 0.85 & 0.20 & $*$ \\
\bottomrule
  \end{tabular}
  \vspace{1mm}
  
  \footnotesize{$*$ indicates that the model failed to produce a valid output.}
\end{table}

\begin{table}[htbp]
  \centering
  \caption{Scalability Analysis: Recall}
  \label{tab:scalability_recall}
  \begin{tabular}{@{}lcccc@{}}
    \toprule
    Total Elements in ACM & \begin{tabular}[c]{@{}c@{}}Gemini Flash\\ (COT)\end{tabular} & \begin{tabular}[c]{@{}c@{}}Gemini Pro\\ (COT)\end{tabular} & \begin{tabular}[c]{@{}c@{}}Gemini Flash\\ (Prompt 1)\end{tabular} & \begin{tabular}[c]{@{}c@{}}Gemini Pro\\ (Prompt 1)\end{tabular} \\
    \midrule
500 & 1.00 & 1.00 & 1.00 & 0.96 \\
700 & 0.99 & 0.98 & 0.54 & 0.92 \\
1000 & 1.00 & 1.00 & 0.98 & 1.00 \\
1500 & 0.81 & 0.90 & 0.98 & 1.00 \\
2000 & 0.96 & 1.00 & 0.60 & 0.97 \\
3000 & 0.68 & 1.00 & 0.12 & 0.41 \\
4000 & 0.82 & 0.85 & 0.93 & 0.63 \\
5000 & 0.15 & 0.56 & 0.27 & $*$ \\
\bottomrule
  \end{tabular}
  \vspace{1mm}
  
  \footnotesize{$*$ indicates that the model failed to produce a valid output.}
\end{table}

The F1 Score (Table~\ref{tab:scalability_f1}), which balances Precision and Recall, confirms this overall trend. The Prompt 1 configurations are clearly unsuitable, with F1 scores often between 0.2 and 0.4 (e.g., 0.21 for Flash (Prompt 1) at 3000 and 0.27 for Pro (Prompt 1) at 4000). The COT models, while capable of perfect results at smaller scales (e.g., 1.00 at 1000 elements), become increasingly unreliable, with sharp drops such as 0.42 for Flash (COT) at 1500 and 0.24 at 5000, or 0.36 for Pro (COT) at 4000.

\begin{table}[htbp]
  \centering
  \caption{Scalability Analysis: F1 Score}
  \label{tab:scalability_f1}
  \begin{tabular}{@{}lcccc@{}}
    \toprule
    Total Elements in ACM & \begin{tabular}[c]{@{}c@{}}Gemini Flash\\ (COT)\end{tabular} & \begin{tabular}[c]{@{}c@{}}Gemini Pro\\ (COT)\end{tabular} & \begin{tabular}[c]{@{}c@{}}Gemini Flash\\ (Prompt 1)\end{tabular} & \begin{tabular}[c]{@{}c@{}}Gemini Pro\\ (Prompt 1)\end{tabular} \\
    \midrule
500 & 0.86 & 0.95 & 0.47 & 0.85 \\
700 & 0.99 & 0.94 & 0.35 & 0.48 \\
1000 & 1.00 & 1.00 & 0.99 & 1.00 \\
1500 & 0.42 & 0.95 & 0.65 & 0.39 \\
2000 & 0.98 & 0.70 & 0.75 & 0.71 \\
3000 & 0.79 & 0.53 & 0.21 & 0.36 \\
4000 & 0.90 & 0.36 & 0.68 & 0.27 \\
5000 & 0.24 & 0.68 & 0.23 & $*$ \\
\bottomrule
  \end{tabular}
  \vspace{1mm}
  
  \footnotesize{$*$ indicates that the model failed to produce a valid output.}
\end{table}

Overall, the results of our comprehensive experiments and their analysis highlight that instability is not absolute, especially at smaller scales. In the majority of test cases up to 2000 elements, at least one model configuration achieves an F1 Score above 0.94, demonstrating the latent capability of LLMs to infer correct policies under controlled conditions. Moreover, COT based prompt consistently performs better than a plain prompt like Prompt 1, indicating policy engineering to be vital for our success. However, as ACM size exceeds 2000 elements, performance deteriorates rapidly. This degradation—manifesting as policy set expansion, severe precision loss, and generation failures—suggests that current LLMs, in the tested configurations, lack the reliability and scalability required for direct application in large-scale, mission-critical ABAC policy mining tasks.

\section{Related Work}
\label{sec:related}



ABAC offers a flexible and fine-grained approach to managing access in complex IT environments \cite{hu2014guide}. Unlike traditional models like RBAC \cite{sandhu1996role}, ABAC grants or denies access based on evaluating policies against attributes of subjects, objects, and the environment. This allows for context-aware rules that adapt dynamically. However, the power of ABAC comes with the significant challenge of policy engineering – the process of defining, managing, and optimizing the potentially large and intricate set of rules required \cite{das2018policy}. Manually creating and maintaining these policies is complex, costly, and error-prone, motivating the search for automated methods. As defined by NIST, this includes the difficulty of translating high-level ``Natural Language Access Control Policies (NLACPs)'' into machine-enforceable "Digital Policies (DPs)" \cite{hu2014guide}. Academic research has thus focused on automating ABAC policy engineering, which is broadly divided into two categories: top-down policy generation and bottom-up policy mining \cite{das2018policy}.


Top-down approaches derive rules from high-level organizational requirements or natural language policies \cite{das2018policy}. While this method can align policies closely with business intent, translating ambiguous natural language into precise, logical rules is a key difficulty. Early research in this area focused on using machine learning to bridge this gap. For instance, Narouei et al. \cite{narouei2017topdown} proposed a top-down framework using deep recurrent neural networks (RNNs) to automatically extract access control policy statements from unrestricted natural language documents. This represented a significant step in automating the translation of NLACPs to-machine-readable formats. This line of research, which explores generation of policies from high-level specifications, has recently been advanced through the use of Large Language Models. More recently, Yang et al. \cite{yang2025llmabac} introduced an LLM-driven knowledge distillation approach to extract machine-enforceable ABAC policies from natural language text, demonstrating the feasibility of automated structured policy translation. In contrast, bottom-up approaches, or policy mining, infer rules by analyzing existing authorization data like access logs or legacy configurations (e.g., ACLs) \cite{das2018policy}. This is the specific problem domain of the present study. The goal is to find a consistent and minimal set of rules that accurately represent observed access patterns. This problem is the conceptual successor to the well-studied field of role mining in RBAC, which sought to discover optimal role sets from user-permission assignments \cite{mitra2016survey}. ABAC policy mining is significantly more complex due to the multi-dimensional nature of attributes.

The foundational work by Xu and Stoller \cite{xu2015mining} introduced one of the first algorithms for mining ABAC policies from ACLs and attribute data. Since then, the field has produced numerous, highly specialized algorithms to tackle this complex optimization problem. For example, recognizing that rule complexity impacts performance, Gautam et al. \cite{gautam_et_al} investigated constrained policy mining to optimize for policies with a limited number of attributes. More recently, Aggarwal and Sural proposed "RanSAM," a randomized search algorithm specifically designed for ABAC policy mining \cite{aggarwal2023ransam}. Other advanced methods include using visual representation and SAT-based model counting to enforce Separation-of-Duty (SOD) constraints \cite{sun2020policy} and employing graph-theoretic methods like biclique analysis \cite{medina2024abac}. Das et al. \cite{das2019vismap} introduced VisMAP, a visual mining approach that enables interactive exploration and optimization of mined ABAC policies, facilitating better understanding of rule structures and dependencies. Additionally, Batra et al. \cite{batra2025semantic} revisited the semantics of ABAC policy mining and enforcement, highlighting that incorrect treatment of object attribute hierarchies can result in misconfigurations, and proposed semantically consistent algorithms to address this issue.
This body of work demonstrates that the state-of-the-art in policy mining relies on customized, computationally complex, and highly specialized algorithms. A key motivation for our study is to investigate whether a general-purpose LLM, without specialized programming, can serve as a viable alternative to these complex algorithmic approaches.


Application of machine learning to access control has also been explored in the last few years. A relatively recent survey on this topic highlights a progression of ML techniques used for tasks like attribute extraction, log analysis, and policy verification \cite{aln-nss-survey2022}. Traditional supervised ML methods, such as the use of Deep Learning to process noisy logs \cite{mocanu2015towards}, were often hampered by a "data scarcity" problem, as large, high-quality, labeled datasets for access control are rare \cite{aln-nss-survey2022, mai2025llm}. LLMs, pre-trained on vast text corpora, possess strong capabilities in natural language understanding, logical reasoning, and pattern recognition, including few-shot learning \cite{mai2025llm}. These abilities suggest their potential to overcome the data scarcity issues and tackle policy engineering tasks. LLMs are already being explored for various cybersecurity applications \cite{mai2025llm}, leading to investigations in the ABAC domain for two primary tasks, namely, (i) Using LLMs to translate high-level requirements into machine-enforceable ABAC rules and (ii) Tasking LLMs with analyzing authorization data to infer underlying ABAC rules. 
An interesting recent work by Mai et al. \cite{mai2025llm} focuses on the first task: policy generation for an Industrial Control System (ICS) network based on NIST guidelines. Their sophisticated framework employs a Mixture-of-Agents (MoA) approach, Retrieval-Augmented Generation (RAG) with a dedicated knowledge base, and structured prompting to generate executable policies \cite{mai2025llm}. Crucially, they found that achieving high accuracy required a mandatory post-processing step: "priority optimization." This step uses ground truth access decisions to mathematically optimize the priority values assigned by the LLMs, indicating that the raw, unassisted LLM output required significant refinement for reliable decision-making \cite{mai2025llm}.


In contrast to all the above work, we investigate the second, complementary task: the capability of standard, contemporary LLMs (Google Gemini and OpenAI ChatGPT) to perform ABAC policy mining directly from authorization data in a synthetic environment. Our research contrasts with Mai et al. \cite{mai2025llm} and other related work in several key aspects, as summarized below.

\begin{description}
  \item[\textbf{Algorithmic Mining}] Specialized algorithms for bottom-up mining from ACLs/logs; computationally heavy and not easily generalizable \cite{gautam_et_al, aggarwal2023ransam}.
  \item[\textbf{ML-based Generation}] RNN-based top-down generation from natural language policies; good textual extraction but poor logical inference \cite{narouei2017topdown}.
  \item[\textbf{LLM-based Generation}] LLMs using MoA and RAG for top-down generation; high accuracy but relies on post-processing \cite{mai2025llm}.
  \item[\textbf{This Study}] Uses standard LLMs with prompt-based bottom-up mining on synthetic authorization data; demonstrates baseline capability, but scalability remains limited.
\end{description}

The primary contrasts are as follows: \begin{description}
    \item \textbf{Task:} Our work focuses on Policy Mining (inductive reasoning from data), whereas Mai et al. \cite{mai2025llm} focus on Policy Generation (deductive reasoning from guidelines). \item \textbf{Methodology:} We benchmark standard, off-the-shelf LLMs using various prompt strategies. This differs from both the complex, bespoke algorithms of Sural et al. \cite{aggarwal2023ransam} and the advanced custom LLM framework (MoA, RAG) of Mai et al. \cite{mai2025llm}. \item \textbf{Evaluation:} The work done in this paper assesses the raw capability of LLMs to produce accurate and minimal rule sets, with a central focus on scalability. This contrasts with Mai et al. \cite{mai2025llm}, who evaluate performance after an essential, ground-truth-dependent optimization step. \end{description}

To summarize, the primary unique contributions of this paper with the respect of the current state of the art are: \begin{enumerate}[i] \item Providing an empirical benchmark of standard LLMs for the direct ABAC policy mining task, establishing baseline capabilities. \item Analyzing the impact of different LLMs, prompt strategies (Zero-Shot, Few-Shot, Chain-of-Thought), and input data formats on mining performance. \item Conducting a critical scalability analysis that reveals significant degradation in LLM performance (accuracy, precision, rule minimality) as the complexity of the ABAC system increases. \end{enumerate}

\section{Conclusion}
\label{sec:concl}

In this paper, we have conducted an empirical investigation into the capabilities of LLMs for the task of policy mining in ABAC systems. Through a phased experimental protocol, we studied the impact of various models, prompting strategies, and data input formats on performance, with the ultimate goal of determining the viability of these models for real-world systems. Our findings indicate that LLMs, particularly Gemini Pro, exhibit a clear potential for this logical inference task, but their effectiveness is highly dependent on the experimental configuration. We determined that both prompt engineering and data representation are critical. Example and Reasoning based prompts (like Prompt 1) and Chain of Thought reasoning significantly outperform naive zero-shot or few-shot prompts. Moreover, COT outperform Prompt 1, highlighting the importance of identifying effective prompt engineering strategies. Furthermore, a flattened `Access Data' log format yield more stable results than structured `ACM' or `ACL' inputs.

However, a scalability analysis revealed the primary limitation of the current approach. As the problem size increases, all the tested configurations demonstrate significant performance degradation. This instability manifested as a failure to maintain rule minimality, catastrophic drops in precision (leading to critical false positives), and an inability to reliably produce any valid output at larger scales. While we observed that in the majority of test cases, at least one configuration could find a high quality solution, the lack of a single, reliable configuration that performs well across all scales is the central challenge. Therefore, we conclude that while LLMs show promise for assisting in policy mining, their performance remains too erratic for direct, autonomous application in large-scale, real world ABAC environments, where accuracy and precision are paramount. The primary issue is not a lack of capability, but rather a lack of reliability.

The limitations identified in this paper suggest several promising directions for future work, moving away from treating the LLM as a solitary black box solver towards a more integrated, guided approach. Our current study relies entirely on the LLM's emergent reasoning to discover the logic and methods for solving the policy mining problem. Future work would focus on hybrid approaches. One such direction is to combine classical, algorithm-based static policy mining methods with LLM-based optimizations. For example, a traditional algorithm could be used to generate a correct, but potentially non-optimized, set of specific rules. The LLM could then be employed for the tasks it excels at, i.e., logical simplification, merging redundant rules, and generalizing the formal policy set into a more minimal, human-readable format.

Another significant avenue is the use of Retrieval-Augmented Generation (RAG). Instead of relying solely on the model's pre-trained knowledge, the prompt would be supplemented with a curated knowledge base. This context could include published state-of-the-art policy mining algorithms, formal principles of ABAC policy design, or organization-specific security requirements. Providing such hints of approach is expected to ground the model's reasoning process, guiding it to follow a proven methodology rather than attempting to reinvent one, which may substantially improve consistency and scalability.

Finally, future systems are likely to implement a human-in-the-loop or formal-verification-in-the-loop process, where an LLM generates a candidate policy set, which is then formally verified by an external tool (such as the evaluation script used in this study). The identified flaws, such as specific false positives or negatives, are then be fed back to the LLM in an iterative refinement loop, allowing the model to correct its own errors. These hybrid and guided strategies represent the most viable path to bridging the gap between the erratic potential and robust reliability required for ABAC policy engineering.

\bibliographystyle{unsrt}
\bibliography{Bibliography}

\appendix
\section{Appendix}
\label{sec:appendix}

In this Appendix, we list the different prompts used in our work and discussed in Sub-section \ref{subsec:prompt_method}.
\subsection{Prompt 1}
\label{subsec:prompt_1}

\begin{quote}

You are an AI specializing in security policy engineering, with expertise in mining 
\textbf{Attribute-Based Access Control (ABAC)} rules from raw access logs.

Your mission is to analyze the provided ABAC dataset and derive a concise and minimal 
set of rules that accurately reflect the authorization logic for \texttt{permit} decisions.

\subsection*{Input Data Format}

The dataset consists of space-separated lines, where each line represents an access request:

\texttt{S\_1\_2 S\_2\_4 ... O\_1\_5 O\_2\_1 ... <decision>}

\begin{itemize}
    \item \textbf{Subject Attributes:} \texttt{S\_i\_j} corresponds to the attribute-value pair 
          (\texttt{SA\_i}, \texttt{S\_i\_j}).
    \item \textbf{Object Attributes:} \texttt{O\_i\_j} corresponds to the attribute-value pair 
          (\texttt{OA\_i}, \texttt{O\_i\_j}).
    \item \textbf{Decision:} 1 for \texttt{permit}, 0 for \texttt{deny}.
\end{itemize}

\subsection*{Tasks}

\begin{itemize}
    \item \textbf{Parse and Filter:} Process the entire dataset and isolate all entries 
          with a \texttt{permit} (1) decision.
    \item \textbf{Generate and Minimize Rules:} Create the smallest possible set of rules 
          that covers all \texttt{permit} entries. A rule can be generalized by identifying 
          a common subset of attributes across multiple permit entries.
    \item \textbf{Critical Safety Constraint:} A generalized rule is only valid if its 
          conditions do not match any \texttt{deny} (0) entries in the original dataset. 
          Your rules must be sound and not grant unintended access. Your rules must also be complete and grant all intended access.
\end{itemize}

\subsection*{Illustrative Example}

Consider this tiny dataset:

\begin{verbatim}
S_1_1 S_2_1 O_1_1 O_2_1 1
S_1_2 S_2_2 O_1_1 O_2_1 1
S_1_3 S_2_1 O_1_2 O_2_1 0
\end{verbatim}

\textbf{Logic:} The two "permit" entries share the exact same object attributes: 
(\texttt{OA\_1}, \texttt{O\_1\_1}) and (\texttt{OA\_2}, \texttt{O\_2\_1}).

This suggests a potential general rule based only on these common attributes:\newline
\{`rule': [(`OA_1', `O_1_1'), (`OA_2', `O_2_1')], `decision': `permit'\}

\textbf{Safety Check:} The deny entry has (\texttt{OA\_1}, \texttt{O\_1\_2}), which does not match the rule's condition (\texttt{OA\_1}, \texttt{O\_1\_1}).

\textbf{Conclusion:} The generalized rule is safe and valid. It correctly covers both permit cases and excludes the deny case.

\subsection*{Final Output Format}

\begin{itemize}
    \item The final output for the main dataset must be a list of generated rules, adhering 
          to these strict formatting requirements:
    \item Each rule must be a single-line Python dictionary string in this exact format:\newline
\{`rule': [(`OA\_1', `O\_1\_7'), (`SA\_2', `S\_2\_2'), ...], `decision': `permit'\}
    \item The output should contain only these rule strings. Do not include any explanations, 
          headings, or other text.
    \item Each rule string must be on a new line.
\end{itemize}
\end{quote}

\subsection{Prompt 2}
\label{subsec:prompt_2}

\begin{quote}
You are an AI specializing in security policy engineering, with an expertise in mining Attribute-Based Access Control (ABAC) rules from raw access logs.
Your mission is to analyze the provided ABAC dataset and derive a concise and minimal set of rules that accurately reflect the authorization logic for ``permit'' decisions.

\textbf{Input Data Format}\newline
The dataset consists of space-separated lines, where each line represents an access request:
S_1_2 S_2_4 ... O_1_5 O_2_1 ... <decision>\newline
Subject Attributes: S_i_j corresponds to the attribute-value pair (`SA_i', `S_i_j').\newline
Object Attributes: O_i_j corresponds to the attribute-value pair (`OA_i', `O_i_j').\newline
Decision: 1 for permit, 0 for deny.

\textbf{Your Tasks}\newline
\textbf{Parse and Filter:} Process the entire dataset and isolate all entries with a ``permit'' (1) decision.\newline
\textbf{Generate and Minimize Rules:} Create the smallest possible set of rules that covers all ``permit'' entries.\newline
A rule can be generalized by identifying a common subset of attributes across multiple permit entries.\newline
\textbf{Critical Safety Constraint:} A generalized rule is only valid if its conditions do not match any ``deny'' (0) entries in the original dataset. Your rules must be sound and not grant unintended access. Your rules must also be complete and grant all intended access.\newline
\textbf{Format the Final Output:} The final output for the main dataset must be a list of generated rules, adhering to these strict formatting requirements:
Each rule must be a single-line Python dictionary string in this exact format:\newline
\{`rule': [(`OA_1', `O_1_7'), (`SA_2', `S_2_2'), ...], `decision': `permit'\}\newline
The output should contain only these rule strings. Do not include any explanations, headings, or other text.
Each rule string must be on a new line.
\end{quote}

\subsection{Prompt 3}
\label{subsec:prompt_3}

\begin{quote}
You are an AI specializing in security policy engineering, with an expertise in mining Attribute-Based Access Control (ABAC) rules from raw access logs.
Your mission is to analyze the provided ABAC dataset and derive a concise and minimal set of rules that accurately reflect the authorization logic for ``permit'' decisions.

\textbf{Input Data Format}\newline
The dataset consists of space-separated lines, where each line represents an access request:
S\_1\_2 S\_2\_4 ... O\_1\_5 O\_2\_1 ... <decision>\newline
Subject Attributes: S\_i\_j corresponds to the attribute-value pair (`SA_i', `S_i_j').\newline
Object Attributes: O\_i\_j corresponds to the attribute-value pair (`OA_i', `O_i_j').\newline
Decision: 1 for permit, 0 for deny.

\textbf{Your Tasks}\newline
\textbf{Parse and Filter:} Process the entire dataset and isolate all entries with a ``permit'' (1) decision.\newline
\textbf{Generate and Minimize Rules:} Create the smallest possible set of rules that covers all ``permit'' entries.\newline
A rule can be generalized by identifying a common subset of attributes across multiple permit entries.\newline
\textbf{Critical Safety Constraint:} A generalized rule is only valid if its conditions do not match any ``deny'' (0) entries in the original dataset. Your rules must be sound and not grant unintended access. Your rules must also be complete and grant all intended access.\newline
\textbf{Format the Final Output:} The final output for the main dataset must be a list of generated rules, adhering to these strict formatting requirements:
Each rule must be a single-line Python dictionary string in this exact format:\newline
\{`rule': [(`OA_1', `O_1_7'), (`SA_2', `S_2_2'), ...], `decision': `permit'\}\newline
The output should contain only these rule strings. Do not include any explanations, headings, or other text.\newline
Each rule string must be on a new line.

\textbf{Example}\newline
To clarify the minimization process, consider this tiny dataset:

Example 1:

S\_1\_1 S\_2\_1 O\_1\_1 O\_2\_1 1\newline
S\_1\_2 S\_2\_2 O\_1\_1 O\_2\_1 1\newline
S\_1\_3 S\_2\_1 O\_1\_2 O\_2\_1 0\newline
Correct Minimal Output for this Example:\newline
\{`rule': [(`OA_1', `O_1_1'), (`OA_2', `O_2_1')], `decision': `permit'\}

Example 2:

S\_1\_1 S\_2\_3 S\_3\_3 O\_1\_7 O\_2\_4 1\newline
S\_1\_1 S\_2\_3 S\_3\_3 O\_1\_3 O\_2\_3 1\newline
S\_1\_1 S\_2\_3 S\_3\_3 O\_1\_3 O\_2\_1 1\newline
...\newline
[Note: Example 2 dataset truncated for brevity.]\newline
Correct Minimal Output for this Example:\newline
\{`rule': [(`OA_1', `O_1_3')], `decision': `permit'\}\newline
\{`rule': [(`SA_1', `S_1_1'), (`SA_2', `S_2_3')], `decision': `permit'\}

Example 3:

S\_1\_1 S\_2\_4 S\_3\_3 S\_4\_3 S\_5\_3 O\_1\_2 O\_2\_3 O\_3\_2 O\_4\_4 O\_5\_1 1\newline
S\_1\_1 S\_2\_4 S\_3\_3 S\_4\_3 S\_5\_3 O\_1\_5 O\_2\_3 O\_3\_3 O\_4\_1 O\_5\_1 0\newline
S\_1\_1 S\_2\_4 S\_3\_3 S\_4\_3 S\_5\_3 O\_1\_3 O\_2\_1 O\_3\_3 O\_4\_4 O\_5\_3 0\newline
...\newline
[Note: Example 3 dataset truncated for brevity.]\newline
Correct Minimal Output for this Example:\newline
\{`rule': [(`OA_2', `O_2_3'), (`OA_4', `O_4_4')], `decision': `permit'\}\newline
\{`rule': [(`OA_5', `O_5_1'), (`SA_2', `S_2_3')], `decision': `permit'\}\newline
\{`rule': [(`OA_1', `O_1_3'), (`OA_3', `O_3_4'), (`SA_3', `S_3_2'), (`SA_4', `S_4_5')], `decision': `permit'\}
\end{quote}

\subsection{Chain of Thoughts Prompt}
\label{subsec:COT_prompt}

\begin{quote}
You are an AI specializing in security policy engineering, with expertise in mining Attribute-Based Access Control (ABAC) rules from raw access logs.

Your mission:
Analyze the provided ABAC dataset and derive a concise and minimal set of rules that accurately reflect the authorization logic for ``permit'' decisions.

\textbf{Input Format}\newline
Each line contains space-separated tokens:\newline
S\_1\_2 S\_2\_4 ... O\_1\_5 O\_2\_1 ... <decision>

\textbf{Where:}
\begin{itemize}
    \item S\_i\_j = subject attribute SA\_i = S\_i\_j
    \item O\_i\_j = object attribute OA\_i = O\_i\_j
    \item decision = 1 for permit, 0 for deny
\end{itemize}

\textbf{Step-by-Step Reasoning (think before answering)}
\begin{enumerate}
    \item \textbf{Parse and Filter:} Separate all entries with decision = 1 (permits) and 0 (denies).
    \item \textbf{Find Common Attribute Patterns:} For each group of permit entries, identify shared subsets of attributes that can generalize them.
    \item \textbf{Validate Generalization:} A general rule is valid only if no deny entry matches all the attributes in the rule.
    \item \textbf{Minimize the Rules:} Combine or simplify rules so that they cover all permit entries without overlap or redundancy.
    \item \textbf{Format the Result:} Output only the minimal valid permit rules, each formatted exactly as:\newline
    \{`rule': [(`OA\_1', `O\_1\_7'), (`SA\_2', `S\_2\_2'), ...], `decision`: `permit'\}
\end{enumerate}

Each new rule must be on a new line. Do not include explanations or extra text.

\textbf{Critical Safety Constraint}\newline
A rule is invalid if it would match any deny (0) entry in the original dataset. Also, the set of all rules should grant all the permit (1).

\textbf{Example}\newline
\textbf{Input Format \quad Decision}\newline
1 \quad S\_1\_2 S\_2\_4 O\_1\_5 O\_2\_1 1\newline
2 \quad S\_1\_2 S\_2\_3 O\_1\_5 O\_2\_1 1\newline
3 \quad S\_1\_2 S\_2\_4 O\_1\_5 O\_2\_2 0\newline
4 \quad S\_1\_1 S\_2\_4 O\_1\_5 O\_2\_1 0

\textbf{Step-by-Step Reasoning (Example View)}\newline
\textbf{Step \quad Example Action \quad Intermediate Result}\newline
1. Parse \& Filter \quad Separate 1, 2 from 3, 4. \quad Permits: 1, 2 / Denies: 3, 4\newline
2. Find Patterns \quad P1 and P2 share: S\_1\_2, O\_1\_5, and O\_2\_1. \quad Draft Rule A: S\_1\_2 $\wedge$ O\_1\_5 $\wedge$ O\_2\_1\newline
3. Validate \quad Check if Draft Rule A matches any Deny. \quad Rule A matches neither D1 nor D2. $\rightarrow$ VALID\newline
4. Minimize \quad Rule A covers both P1 and P2 with 3 attributes. (Minimal) \quad Final Rule Set: \{Rule A\}

\textbf{Final Formatted Output:}\newline
\{`rule': [(`SA\_1', `S\_1\_2'), (`OA\_1', `O\_1\_5'), (`OA\_2', `O\_2\_1')], `decision': `permit'\}

\textbf{Output}\newline
Only output the final list of minimal safe rules. Each new rule must be on a new line. No explanations, no comments, just the rule dictionaries.
\end{quote}

\subsection{No 0 to 1 Prompt}
\label{subsec:no_0_to_1_prompt}

\begin{quote}
You are an AI specializing in security policy engineering, with an expertise in mining Attribute-Based Access Control (ABAC) rules from raw access logs.

Your mission is to analyze the provided ABAC dataset and derive a concise and minimal set of rules that accurately reflect the authorization logic for "permit" decisions.

\textbf{Input Data Format}\newline
The dataset consists of space-separated lines, where each line represents an access request: \newline
\texttt{S\_1\_2 S\_2\_4 ... O\_1\_5 O\_2\_1 ... <decision>}

\textbf{Subject Attributes:} \texttt{S\_i\_j} corresponds to the attribute-value pair (\texttt{`SA\_i'}, \texttt{`S\_i\_j'}).

\textbf{Object Attributes:} \texttt{O\_i\_j} corresponds to the attribute-value pair (\texttt{`OA\_i'}, \texttt{`O\_i\_j'}).

\textbf{Decision:} 1 for permit, 0 for deny.

\textbf{Your Tasks}\newline
\textbf{Parse and Filter:} Process the entire dataset and isolate all entries with a "permit" (1) decision.

\textbf{Generate and Minimize Rules:} Create the smallest possible set of rules that covers all "permit" entries.

A rule can be generalized by identifying a common subset of attributes across multiple permit entries.

Your primary goal is to minimize the number of rules. This minimization process is governed by one absolute safety requirement:

\textbf{Strictly Forbidden (Safety):} A generated rule must never `permit' an access request that is marked as `deny' (0) in the original dataset. Any rule that generalizes in a way that covers even one 'deny' entry is invalid. This prevents "false positives."

\textbf{Acceptable Trade-off (Minimization):} Your generalized rules may be stricter than the original logs. This means a `permit' (1) entry from the original dataset might not be covered by your final rule set (i.e., it is effectively denied). This "false negative" is an acceptable cost only if it allows for a simpler, more generalized rule set.

\textbf{Illustrative Example}\newline
To clarify the minimization process, consider this tiny dataset:

\textbf{Sample Data:}
\begin{verbatim}
S_1_1 S_2_1 O_1_1 O_2_1 1
S_1_2 S_2_2 O_1_1 O_2_1 1
S_1_3 S_2_1 O_1_2 O_2_1 0
\end{verbatim}

\textbf{Logic:}\newline
The two "permit" entries share the exact same object attributes: (\texttt{`OA\_1'}, \texttt{`O\_1\_1'}) and (\texttt{`OA\_2'}, \texttt{`O\_2\_1'}).

This suggests a potential general rule based only on these common attributes: \texttt{\{`rule': [(`OA\_1', `O\_1\_1'), (`OA\_2', `O\_2\_1')], `decision': `permit'\}}.

\textbf{Safety Check:} We check if this general rule accidentally covers the "deny" entry. The deny entry has (\texttt{`OA\_1'}, \texttt{`O\_1\_2'}), which does not match the rule's condition (\texttt{`OA\_1'}, \texttt{`O\_1\_1'}).

\textbf{Conclusion:} The generalized rule is safe and valid. It correctly covers both permit cases and excludes the deny case.
\textbf{Correct Minimal Output for this Example:}\newline
\texttt{\{`rule': [(`OA\_1', `O\_1\_1'), (`OA\_2', `O\_2\_1')], `decision': `permit'\}}

\textbf{Format the Final Output:} The final output for the main dataset must be a list of generated rules, adhering to these strict formatting requirements:
Each rule must be a single-line Python dictionary string in this exact format: \newline
\texttt{\{`rule': [(`OA\_1', `O\_1\_7'), (`SA\_2', `S\_2\_2'), ...], `decision': `permit'\}}

NO RULE SHOULD GRANT A PERMIT WHICH IS ORIGINALY A DENY.

The output should contain only these rule strings. Do not include any explanations, headings, or other text.

Each rule string must be on a new line.
\end{quote}

\subsection{Deny Allowed Prompt}
\label{subsec:deny_prompt}
The rule set operates on a \textbf{Deny-Overrides Principle}. This is the most important concept for your task.

When an access request is evaluated:
\begin{itemize}
    \item It is checked against all rules in the set.
    \item If the request matches one or more `deny' rules, the final decision is \textbf{deny}. This is true even if it also matches `permit' rules.
    \item If the request matches no `deny' rules but matches one or more `permit' rules, the final decision is \textbf{permit}.
    \item If a request matches no rules at all, the implicit decision is `deny'. (Your generated rule set must ensure every entry in the dataset is correctly classified by this logic).
\end{itemize}

\textbf{Input Data}\newline
The dataset consists of space-separated lines, where each line represents an access request:
\begin{verbatim}
S_1_2 S_2_4 ... O_1_5 O_2_1 ... <decision>
\end{verbatim}

\begin{itemize}
    \item \textbf{Subject Attributes:} \texttt{S\_i\_j} corresponds to the attribute-value pair \texttt{(`SA\_i', `S\_i\_j')}.
    \item \textbf{Object Attributes:} \texttt{O\_i\_j} corresponds to the attribute-value pair \texttt{(`OA\_i', `O\_i\_j')}.
    \item \textbf{Decision:} \texttt{1} for permit, \texttt{0} for deny.
\end{itemize}

\begin{itemize}
    \item \textbf{Parse and Classify:} Process the entire dataset and separate the entries into two groups: "permit" (1) and "deny" (0).
    \item \textbf{Generate and Minimize Rules:} Create the smallest possible set of rules that correctly classifies every entry in the dataset, using the Deny-Overrides Principle.
    \item \textbf{Minimization Strategy:} Your goal is to find the rule set \textit{R} with the minimum number of rules.
    
    This strategy requires you to find conflicts and resolve them. For example, you could create broad, general `permit' rules that cover many permit entries.
    
    If a broad `permit' rule accidentally covers a `deny' entry, you must then create a more specific `deny' rule that acts as an exception.
    
    Conversely, you could create broad `deny' rules and add specific `permit' rules as exceptions.
    
    You must choose the strategy (e.g., broad-permit-with-specific-deny vs. broad-deny-with-specific-permit) that results in the absolute smallest total number of rules.
\end{itemize}

\textbf{Illustrative Example}\newline
This example demonstrates how to use the Deny-Overrides principle for minimization.

\textbf{Sample Data:}
\begin{verbatim}
S_1_1 S_2_1 O_1_1 O_2_1 1 (Permit 1)
S_1_2 S_2_2 O_1_1 O_2_1 1 (Permit 2)
S_1_3 S_2_1 O_1_1 O_2_1 0 (Deny 1)
S_1_4 S_2_1 O_1_1 O_2_1 1 (Permit 3)
\end{verbatim}

\textbf{Logic:}
\begin{itemize}
    \item All four entries share the attributes \texttt{(`OA\_1', `O\_1\_1')} and \texttt{(`OA\_2', `O\_2\_1')}.
    \item A naive approach would create 4 specific rules (one for each line), resulting in a set of 4 rules. This is not minimal.
    \item A minimal approach uses the Deny-Overrides principle:
\end{itemize}

\begin{description}
    \item[Step 1:] Create a single, broad `permit' rule to cover all permit cases. A good candidate is:
    \begin{verbatim}
{`rule': [(`OA_1', `O_1_1'), (`OA_2', `O_2_1')], `decision': `permit'}
    \end{verbatim}
    \item[Step 2:] Check this rule. It correctly covers `Permit 1', `Permit 2', and `Permit 3'. However, it \textit{incorrectly} covers `Deny 1' (which would be permitted).
    \item[Step 3:] Create a specific `deny' rule to act as an exception for the `Deny 1' case. This rule must be specific enough to only match `Deny 1'.
    \begin{verbatim}
{`rule': [(`SA_1', `S_1_3'), (`SA_2', `S_2_1'), 
 (`OA_1', `O_1_1'), (`OA_2', `O_2_1')], `decision': `deny'}
    \end{verbatim}
    (Note: A less specific rule like \texttt{[(`SA\_1', `S\_1\_3'), (`OA\_1', `O\_1\_1')]} would also work if it doesn't conflict with other permit entries not shown in this small sample.)
\end{description}

\textbf{Minimal Rule Set (2 rules):}
\begin{verbatim}
{`rule': [(`OA_1', `O_1_1'), (`OA_2', `O_2_1')], `decision': `permit'}
{`rule': [(`SA_1', `S_1_3'), (`SA_2', `S_2_1'), 
 (`OA_1', `O_1_1'), (`OA_2', `O_2_1')], `decision': `deny'}
\end{verbatim}

\textbf{Evaluation Check:}
\begin{itemize}
    \item `Permit 1' (S\_1\_1...): Matches `permit' rule, does not match `deny' rule. -> \textbf{Permit}. (Correct)
    \item `Deny 1' (S\_1\_3...): Matches `permit' rule AND `deny' rule. -> Deny-Overrides applies. -> \textbf{Deny}. (Correct)
\end{itemize}

\textbf{Format the Final Output}\newline
The final output must be a list of generated rules, adhering to these strict formatting requirements:
\begin{itemize}
    \item Each rule must be a single-line Python dictionary string in this exact format:
\begin{verbatim}
{`rule': [(`OA_1', `O_1_7'), (`SA_2', `S_2_2'), ...],
`decision': `permit'}
\end{verbatim} OR
\begin{verbatim}
{`rule': [(`OA_1', `O_1_7'), (`SA_2', `S_2_2'), ...],
`decision': `deny'}
\end{verbatim}
    \item The output should contain \textbf{only} these rule strings.
    \item Do not include any explanations, headings, or other text.
    \item Each rule string must be on a new line.
\end{itemize}

\subsection{ACM prompt}
\label{subsec:ACM_prompt}

\begin{quote}
You are an AI assistant specializing in security policy engineering. Your expertise is in mining minimal, sound Attribute-Based Access Control (ABAC) policies from access logs and attribute data.

\textbf{Objective:} Your mission is to analyze the provided data file, identify all "permit" and "deny" access patterns, and generate the most concise (minimal) set of ABAC rules that perfectly recreates all "permit" decisions without ever incorrectly authorizing a "deny" decision.

\textbf{Input Data Specification}\newline
You will be given two files:

\textbf{1. Attribute Data (output.json)}\newline
This JSON file contains attribute values for all subjects and objects.
\begin{itemize}
    \item \texttt{"SV"}: A list of lists. \texttt{SV[i]} is the attribute list for Subject \texttt{i}.
    \item \texttt{"OV"}: A list of lists. \texttt{OV[i]} is the attribute list for Object \texttt{i}.
\end{itemize}

\textbf{Attribute Naming Convention:} Attribute names are derived from their 0-based index in the attribute lists.
\begin{itemize}
    \item The \texttt{k}-th element in an \texttt{SV} list (\texttt{SV[i][k]}) corresponds to the attribute name $SA_{k+1}$.
    \item The \texttt{k}-th element in an \texttt{OV} list (\texttt{OV[i][k]}) corresponds to the attribute name $OA_{k+1}$.
\end{itemize}

\textbf{2. Access Control Matrix (ACM.txt)}\newline
This text file defines all permit and deny decisions using a matrix.

\textbf{Format:} Each row \texttt{i} corresponds to Subject \texttt{i}, and each column \texttt{j} corresponds to Object \texttt{j}.

\textbf{Permit Logic:} A pair (Subject \texttt{i}, Object \texttt{j}) is a permit decision if the value at \texttt{ACM[i][j]} is \texttt{`1'}.

\textbf{Deny Logic:} A pair (Subject \texttt{i}, Object \texttt{j}) is an explicit deny decision if the value at \texttt{ACM[i][j]} is \texttt{`0'}.

\textbf{Your Precise Workflow}\newline
\textbf{1. Data Reconstruction}
\begin{itemize}
    \item Load all subject attributes from \texttt{"SV"} and object attributes from \texttt{"OV"}.
    \item Parse \texttt{ACM.txt} to identify all explicit permit pairs \texttt{(i, j)} (value \texttt{`1'}) and all explicit deny pairs \texttt{(i, j)} (value \texttt{`0'}).
    \item Create two master sets based on these pairs by retrieving their corresponding attributes:
    \item \textbf{PermitSet:} A set of all \texttt{(subject\_attributes, object\_attributes)} tuples that correspond to a "permit" decision (\texttt{`1'}).
    \item \textbf{DenySet:} A set of all \texttt{(subject\_attributes, object\_attributes)} tuples that correspond to a "deny" decision (\texttt{`0'}).
\end{itemize}

\textbf{2. Rule Generation and Minimization}
\begin{itemize}
    \item Your goal is to find the smallest set of rules that covers every entry in the \textbf{PermitSet}.
    \item A rule is a subset of attributes (e.g., \texttt{[(`SA\_1', `S\_1\_3'), (`OA\_3', `O\_3\_1')]}).
    \item A rule \textit{covers} an entry if all attributes in the rule match the corresponding attributes in the entry.
    \item Start by creating the most specific rules (one rule for each unique entry in \textbf{PermitSet}, grouped by their combined attributes).
    \item Iteratively generalize these rules by finding common attribute subsets.
\end{itemize}

\textbf{3. Critical Safety Constraint (Soundness Check)}
\begin{itemize}
    \item This is the most important step.
    \item A generalized rule is \textbf{VALID} only if it does not cover \textit{any} entry in the \textbf{DenySet}.
    \item You must discard any potential generalization, no matter how minimal, if it would incorrectly grant access to a "deny" case.
    \item Your final rules must be 100\% sound and complete.
\end{itemize}

\textbf{Illustrative Example}\newline
\textbf{ACM.txt}
\begin{verbatim}
0 0 0 0
0 0 0 0
1 1 1 1
1 1 1 1
\end{verbatim}

\textbf{output.json}
\begin{verbatim}
{
    "SV": [
        [
            "S_1_1",
            "S_2_3",
            "S_3_2"
        ],
        [
            "S_1_1",
...
[Truncated for brevity]
    "OV": [
        [
            "O_1_5",
            "O_2_4",
            "O_3_1"
        ],
        [
            "O_1_3", 
...
[Truncated for brevity.]
}
\end{verbatim}

\textbf{Analysis:}
\begin{itemize}
    \item \textbf{Permit Pairs:} (S2, O0), (S2, O1), (S2, O2), (S2, O3), (S3, O0), (S3, O1), (S3, O2), (S3, O3).
    \item \textbf{Deny Pairs:} All pairs with S0 or S1, e.g., (S0, O0), (S0, O1), (S1, O2), etc.
    \item \textbf{Permit Attributes:} The attributes for all "permit" subjects (S2, S3) are:
    \item S2: [SA\_1: "S\_1\_3", SA\_2: "S\_2\_2", SA\_3: "S\_3\_5"]
    \item S3: [SA\_1: "S\_1\_3", SA\_2: "S\_2\_1", SA\_3: "S\_3\_3"]
    \item The common attribute between all "permit" subjects is (SA\_1, \texttt{`S\_1\_3'}).
    \item \textbf{Minimization:}
    \item A potential minimal rule is \texttt{\{`rule': [(`SA\_1', `S\_1\_3')], `decision': `permit'\}}.
    \item \textbf{Soundness Check:}
    \item We must check this rule against all Deny Cases.
    \item The "deny" subjects are S0 and S1.
    \item \textbf{S0 Attributes:} [SA\_1: "S\_1\_1", ...] (Rule does not match)
    \item \textbf{S1 Attributes:} [SA\_1: "S\_1\_1", ...] (Rule does not match)
    \item Since the rule (SA\_1, \texttt{`S\_1\_3'}) does not match any "deny" subjects, it is sound and valid.
\end{itemize}

\textbf{Correct Minimal Output for this Example:} \texttt{\{`rule': [(`SA\_1', `S\_1\_3')], `decision': `permit'\}}

\textbf{Final Output Format}\newline
The final output must be only the generated rule strings.
Do not include any explanations, headings, or other text.
Each rule must be a single-line Python dictionary string in this exact format: \newline
\texttt{\{`rule': [(`OA\_1', `O\_1\_7'), (`SA\_2', `S\_2\_2'), ...], `decision': `permit'\}}
\end{quote}

\subsection{ACL prompt}
\label{subsec:ACL_prompt}
\begin{quote}
You are an AI assistant specializing in security policy engineering. Your expertise is in mining minimal, sound Attribute-Based Access Control (ABAC) policies from access logs and attribute data.

\textbf{Objective:} Your mission is to analyze the provided data files, identify all "permit" and "deny" access patterns, and generate the most concise (minimal) set of ABAC rules that perfectly recreates all "permit" decisions without ever incorrectly authorizing a "deny" decision.

\textbf{Input Data Specification}\newline
You will be given two files:

\textbf{1. Attribute Data (output.json)}\newline
This JSON file contains attribute values for all subjects and objects.
\begin{itemize}
    \item \texttt{"SV"}: A list of lists. \texttt{SV[i]} is the attribute list for Subject \texttt{i}.
    \item \texttt{"OV"}: A list of lists. \texttt{OV[i]} is the attribute list for Object \texttt{i}.
\end{itemize}

\textbf{Attribute Naming Convention:} Attribute names are derived from their 0-based index in the attribute lists.
\begin{itemize}
    \item The \texttt{k}-th element in an \texttt{SV} list (\texttt{SV[i][k]}) corresponds to the attribute name $SA_{k+1}$.
    \item The \texttt{k}-th element in an \texttt{OV} list (\texttt{OV[i][k]}) corresponds to the attribute name $OA_{k+1}$.
\end{itemize}

\textbf{2. Access Control List (ACL.txt)}\newline
This file defines all permit decisions using an object-centric adjacency list.

\textbf{Format:} Each line is \texttt{object\_index: subject\_index\_1 subject\_index\_2 ...}

\textbf{Permit Logic:} A pair (Subject \texttt{i}, Object \texttt{j}) is a permit decision if subject \texttt{i} is present in the list for object \texttt{j}.

\textbf{Deny Logic:} A pair (Subject \texttt{i}, Object \texttt{j}) is an implicit deny decision if subject \texttt{i} is \textit{not} present in the list for object \texttt{j}.

\textbf{You must follow this precise workflow:}\newline
\textbf{1. Data Reconstruction}
\begin{itemize}
    \item Load all subject attributes from \texttt{"SV"} and object attributes from \texttt{"OV"}.
    \item Parse \texttt{ACL.txt} to identify all explicit permit pairs \texttt{(i, j)}.
    \item Generate the complete set of implicit deny pairs by iterating through all possible (Subject \texttt{i}, Object \texttt{j}) combinations (where $i < \text{len(SV)}$ and $j < \text{len(OV)}$) and finding those not in the permit list.
    \item Create two master sets based on these pairs:
    \item \textbf{PermitSet:} A set of all \texttt{(subject\_attributes, object\_attributes)} tuples that correspond to a "permit" decision.
    \item \textbf{DenySet:} A set of all \texttt{(subject\_attributes, object\_attributes)} tuples that correspond to a "deny" decision.
\end{itemize}

\textbf{2. Rule Generation and Minimization}
\begin{itemize}
    \item Your goal is to find the smallest set of rules that covers every entry in the \textbf{PermitSet}.
    \item A rule is a subset of attributes (e.g., \texttt{[(`SA\_1', `S\_1\_3'), (`OA\_3', `O\_3\_1')]}).
    \item A rule \textit{covers} an entry if all attributes in the rule match the corresponding attributes in the entry.
    \item Start by creating the most specific rules (one rule for each unique entry in \textbf{PermitSet}).
    \item Iteratively generalize these rules by finding common attribute subsets (e.g., merging two rules that only differ by one attribute into a new rule that omits that attribute).
\end{itemize}

\textbf{3. Critical Safety Constraint (Soundness Check)}\newline
This is the most important step.
\begin{itemize}
    \item A generalized rule is \textbf{VALID} only if it does not cover \textit{any} entry in the \textbf{DenySet}.
    \item You must discard any potential generalization, no matter how minimal, if it would incorrectly grant access to a "deny" case.
    \item Your final rules must be 100\% sound and complete.
\end{itemize}

\textbf{Illustrative Example}\newline
\textbf{ACL.txt}
\begin{verbatim}
0: 2 3
1: 2 3
2: 2 3
3: 2 3
\end{verbatim}

\textbf{output.json}
\begin{verbatim}
{
    "SV": [
        [
            "S_1_1",
            "S_2_3",
            "S_3_2"
        ],
        [
            "S_1_1",
...
[Truncated for brevity]
    "OV": [
        [
            "O_1_5",
            "O_2_4",
            "O_3_1"
        ],
        [
            "O_1_3", 
...
[Truncated for brevity.]
}
\end{verbatim}

\textbf{Analysis:}
\begin{itemize}
    \item \textbf{Permit Pairs:} (S2, O0), (S3, O0), (S2, O1), (S3, O1), (S2, O2), (S3, O2), (S2, O3), (S3, O3).
    \item \textbf{Deny Pairs:} All other combinations, e.g., (S0, O0), (S0, O1), (S1, O2), etc.
    \item \textbf{Permit Attributes:} The attributes for all "permit" subjects (S2, S3) are:
    \item S2: [SA\_1: "S\_1\_3", SA\_2: "S\_2\_2", SA\_3: "S\_3\_5"]
    \item S3: [SA\_1: "S\_1\_3", SA\_2: "S\_2\_1", SA\_3: "S\_3\_3"]
    \item The common attribute between all "permit" subjects is (\texttt{`SA\_1'}, \texttt{`S\_1\_3'}).
    \item \textbf{Minimization:} A potential minimal rule is \texttt{\{`rule': [(`SA\_1', `S\_1\_3')], `decision': `permit'\}}.
    \item \textbf{Soundness Check:} We must check this rule against all Deny Cases.
    \item The "deny" subjects are S0 and S1.
    \item \textbf{S0 Attributes:} [SA\_1: "S\_1\_1", ...] (Rule does not match)
    \item \textbf{S1 Attributes:} [SA\_1: "S\_1\_1", ...] (Rule does not match)
    \item Since the rule (\texttt{`SA\_1'}, \texttt{`S\_1\_3'}) does not match any "deny" subjects, it is sound and valid.
\end{itemize}

\textbf{Correct Minimal Output for this Example:}\newline
\texttt{\{`rule': [(`SA\_1', `S\_1\_3')], `decision': `permit'\}}

\textbf{Final Output Format}\newline
The final output must be only the generated rule strings.
Do not include any explanations, headings, or other text.
Each rule must be a single-line Python dictionary string in this exact format:\newline
\texttt{\{`rule': [(`OA\_1', `O\_1\_7'), (`SA\_2', `S\_2\_2'), ...], `decision': `permit'\}}
\end{quote}

\end{document}